\documentclass{emulateapj}
\usepackage{aas_macros}
\usepackage{amssymb}
\usepackage{natbib}
\usepackage[svgnames]{xcolor}
\makeatletter

\newcommand{\Rmnum}[1]{\expandafter\@slowromancap\romannumeral #1@}
\makeatother

\newcommand  \CII{\,C\,{\footnotesize II}}

\def\mean#1{\left< #1 \right>}

\newcommand  \HII{\,H\,{\footnotesize II}}
\newcommand  \HI{\,H\,{\footnotesize I}}
\newcommand  \m{\mathrm}

\begin{document}

\title{The location, clustering, and propagation of massive star formation \\ in giant molecular clouds}
\author{Bram B. Ochsendorf\altaffilmark{1}, Margaret Meixner\altaffilmark{2,1}, J\'{e}r\'{e}my Chastenet\altaffilmark{2,3}, Alexander G.\,G.\,M. Tielens\altaffilmark{4} \& Julia Roman-Duval\altaffilmark{2}}
\affil{$^1$ Department of Physics and Astronomy, The Johns Hopkins University, 3400 North Charles Street, Baltimore, MD 21218, USA\\ $^2$ Space Telescope Science Institute, 3700 San Martin Drive, Baltimore, MD 21218, USA \\ $^3$ Observatoire Astronomique de Strasbourg, Universit\'{e} de Strasbourg, CNRS, UMR 7550, 11 rue de l'Universit\'{e}, F-67000 Strasbourg, France
 \\ $^4$ Leiden Observatory, Leiden University, P.O. Box 9513, NL-2300 RA, The Netherlands}
\email{bochsen1@jhu.edu}

\begin{abstract}
Massive stars are key players in the evolution of galaxies, yet their formation pathway remains unclear. In this work, we use data from several galaxy-wide surveys to build an unbiased dataset of $\sim$\,700 massive young stellar objects (MYSOs), $\sim$\,200 giant molecular clouds (GMCs), and $\sim$\,100 young ($\textless$\,10 Myr) optical stellar clusters (SCs) in the Large Magellanic Cloud. We employ this data to quantitatively study the location and clustering of massive star formation and its relation to the internal structure of GMCs. We reveal that massive stars do not typically form at the highest column densities nor centers of their parent GMCs at the $\sim$\,6 pc resolution of our observations. Massive star formation clusters over multiple generations and on size scales much smaller than the size of the parent GMC. We find that massive star formation is significantly boosted in clouds near SCs. Yet, whether a cloud is associated with a SC does not depend on either the cloud's mass or global surface density. These results reveal a connection between different generations of massive stars on timescales up to 10 Myr. We compare our work with Galactic studies and discuss our findings in terms of GMC collapse, triggered star formation, and a potential dichotomy between low- and high-mass star formation.
\end{abstract}

\section{Introduction}

Massive stars dominate the structure and energy budget of the interstellar medium of galaxies through intense radiation fields, stellar winds, and supernova explosions. Yet, the pathway that leads to their formation remains unclear, as the process is notoriously difficult to probe because of large distances, crowding, high levels of obscuration, and short lifetimes. In general, star formation studies have seen dramatic progress in the past decade, which can largely be attributed to the {\em Spitzer} space telescope and the {\em Herschel} space observatory. These missions opened up the mid-to-far infrared (IR) sky at high resolution, allowing us to peek into star forming cradles that are deeply embedded within giant molecular clouds (GMCs). 

The internal structure of GMCs reveal infrared dark clouds (IRDCs) and filaments (up to tens of pc), clumps ($\sim$\,1 pc) and cores ($\sim$\,0.1 pc). It is now largely understood that there is an intimate connection between filaments and the formation of low-mass prestellar cores \citep{konyves_2010,andre_2010,andre_2014}. In contrast, studying high-mass clumps and cores has proven to be difficult despite numerous attempts targeting the earliest stages of massive star formation \citep{motte_2007,tackenberg_2012,ragan_2012, schneider_2012}. Recent large surveys of the Galactic plane yield promising results by detecting {\em candidate} massive star forming clumps \citep[e.g.,][]{svoboda_2015}. Still, confusion and distance ambiguity will inherently complicate studies of massive star formation in the Galaxy and its connection to larger-scale structures in the interstellar medium, e.g., the parent GMCs. Leaving aside the difficulties in probing Galactic massive star formation, there is no theoretical consensus as to the exact physical process that ultimately leads to a (cluster of) massive stars \citep{tan_2014}. In this respect, it has long been debated that low-mass stars and high-mass stars may not form alike: whereas low-mass cores and stars may form `spontaneously' through hierarchical fragmentation within GMCs \citep{andre_2014}, the formation of high-mass stars may be `triggered' \citep{elmegreen_1998} by an external mechanism, although the exact nature and/or importance of triggering has remained controversial \citep[see][and references therein]{dale_2015}. 

In this work, we present a galaxy-wide study of massive star formation and its relation with GMCs in the Large Magellanic Cloud (LMC). The LMC provides us with an excellent opportunity to study the formation of massive stars in a wide range of evolutionary stages, since its face-on orientation minimizes confusion and distance ambiguities, while being close enough to resolve individual clouds and stars ($\sim$\,50 kpc; \citealt{pietrzynski_2013}). By combining several galaxy-wide surveys, we create a unique view of massive young stellar objects (MYSOs), GMCs, and optical stellar clusters (SCs) in the LMC. The multi-facetted nature and sheer size of the data traces massive star formation as a function of environment and evolutionary state, and the overarching goal of this study is to exploit this unique dataset to quantify the location, clustering, and propagation of massive star formation within GMCs. In Sec. \ref{sec:observations}, we present the observations. In Sec. \ref{sec:catalogue}, we build our catalogue of MYSOs, the completeness of which is tested in Sec. \ref{sec:completeness}. We describe the dust fitting and creation of column density maps and subsequent cloud decomposition in Sec. \ref{sec:mapandcloud}. The distribution of MYSOs within GMCs and its relation to SCs is presented in Sec. \ref{sec:results}. We compare our results with studies performed in the Galaxy, an discuss our findings in relation to recent numerical and analytical studies of collapsing molecular clouds in Sec. \ref{sec:discussion}. We conclude in Sec. \ref{sec:conclusions}.

\section{Observations}\label{sec:observations}

In this work, we make use of the far-IR images from the Herschel Inventory of the Agents of Galaxy Evolution (HERITAGE; \citealt{meixner_2013}) covering the entire IR-emitting part of the LMC at 70 $\mu$m, 160 $\mu$m, 250 $\mu$m, 350 $\mu$m and 500 $\mu$m at $\sim$\,7", 12", 18", 25", and 36" resolution. In addition, we employ data from the Magellanic Mopra Assessment (MAGMA; \citealt{wong_2011}) Data Release 3 (Wong et al., in prep), a 45"-resolution targeted study of GMCs ($\sim$\,200 in total; Sec. \ref{sec:mapandcloud}) with fluxes greater than 1.2\,$\times$\,10$^5$ K km s$^{-1}$ arcsec$^{2}$ and a completeness limit of $M$\,$\sim$\,3\,$\times$\,10$^4$ M$_\m{\odot}$.

\section{Catalogue of massive young stellar objects}\label{sec:catalogue}

We have compiled a catalog of (highly) probable YSOs by combining the results of galaxy-wide searches of YSO candidates \citep{whitney_2008,gruendl_2009} using Spitzer's Surveying the Agents of a Galaxy's Evolution (SAGE; \citealt{meixner_2006}) data and HERITAGE data \citep{seale_2014}. These works produced YSO catalogues through careful selection criteria (e.g., color-magnitude cuts, morphological inspection) tailored to minimize contamination from sources such as planetary nebulae, evolved stars, and background galaxies. A certain level of contamination is still expected, with estimates ranging from 55\% \citep{whitney_2008}, 20\,-\,30\% \citep{gruendl_2009}, and $\textless$\,10\% \citep{seale_2014}. However, these levels mainly apply to the faint end of the YSO distribution, which overlap more with the aforementioned contaminants in color-magnitude space than their luminous (i.e., higher-mass) counterparts. For the MYSOs, which are the focus of this study, we expect contamination of our YSO catalogues to be less important.

From \citet{whitney_2008}, we use the `YSO candidate' and `high-probability YSO candidate' lists (989 sources). From \citet{gruendl_2009} we restrict ourselves to the `probable' and `definite' class of YSO candidates (1171 sources), and from \citet{seale_2014} we employ the list of 2493 `probable' YSOs. These catalogues inherently have overlapping sources, and thus we throw out duplicates by cross-matching the catalogues by finding the nearest on-sky matches between coordinates. We define a match between the SAGE catalogues as coordinates that are within $\leq$\,2" from one another, while this threshold is raised to $\leq$\,10" for cross-matching the SAGE and HERITAGE catalogues because of the coarser resolution of the HERITAGE photometric bands (up to $\sim$\,36" for the 500 $\mu$m band; \citealt{meixner_2013}). We end up with a final list of 3524 high-probable YSO candidates for the entire LMC. 

We also consider the far-infrared `dust clumps' (DCs) discussed in \citet{seale_2014}. These dust clumps differ from a HERITAGE candidate YSO through a lack of a 24 $\mu$m point-source detection, commonly thought as a robust tracer of star formation \citep{dunham_2014}. \citet{seale_2014} did not include DCs in their final YSO candidate list, as the authors revealed that the photometry of these sources could not distinguish between a highly-embedded YSO, or a starless ISM clump illuminated by a moderate external interstellar radiation field. However, DCs brighter than $L$ $\geq$ 10$^3$ $L_\m{\odot}$ are more luminous than can be explained by the typical radiation field pervading the LMC \citep{seale_2014}, implying the presence of an embedded heating source. Upon closer inspection, \citet{seale_2014} noted that many of the bright DCs reveal extended or saturated 24 $\mu$m emission, preventing their detection as a point-source in this band and, consequently, eluded classification as an YSO in the SAGE catalog of \citet{whitney_2008}. The photometric extraction method employed by \citet{gruendl_2009} did allow for extended objects to enter the catalog. In the remainder of this work, we opted to present the results for the DCs separately from that of the MYSOs, however many DCs may in fact represent true MYSOs due to a 24 $\mu$m misclassification.

In order to characterize the sources within our catalogue, we attempt to fit all sources with the \citet{robitaille_2006} YSO models (except for the DCs, for which there are no suitable models available). The \citet{robitaille_2006} models (2\,$\times$\,10$^5$ in total) cover a wide range of physical parameters for different stages in the YSO evolutionary path, often divided in Stage 1 (least evolved), 2, and 3 (most evolved); see \citet{robitaille_2006} for a definition of these classes. However, it is important to note that the parameters used to divide YSOs in these stages (such as disk mass, envelope accretion, and mass of the central source) are not constrained from the SED alone (for a thorough discussion, see \citealt{robitaille_2008}). The bolometric luminosity of the sources is well constrained by the fits, but given that the source luminosity is expected to evolve during the early stages of star formation (e.g., mass accretion), one cannot simply translate observed luminosity into a (main sequence) mass. Therefore, for the remainder of this study we chose to define our completeness limits (Sec. \ref{sec:completeness}) and the subsequent source analysis in terms of mass predicted by the \citet{robitaille_2006} models, however we caution the reader that the reported masses rely on the accuracy of the integrated evolutionary tracks \citep{robitaille_2008}. 

We define a `well-fitted' source by one yielding a reduced chi square of $\chi^2_\m{red}$ $\leq$ 5. Although arbitrarily chosen, this `chi-by-eye' threshold provides a collection of fits that seem very acceptable. From all 3524 YSO sources, 2558 sources have sufficient photometric constraints to be passed to the \citet{robitaille_2006} SED fitter. From these, 1278 sources have $\chi^2_\m{red}$ $\leq$ 5, out of which 691 are above our completeness limit (i.e., $M$\,$\geq$\,8 $L_\m{\odot}$; Sec. \ref{sec:completeness}). From these 691 MYSOs, we find that 569 are Stage I, 103 are Stage II, and 19 are Stage III. As discussed above, this classification scheme uses parameters not directly related to the SED, but it does provide a handle on the evolutionary state of the source and thereby its age. For low-mass YSOs (Log($L$/$L_\odot$)\,$\lesssim$\,0.5, $M$\,$\sim$\,0.5 M$_\odot$), the estimated lifetimes of Stage I and Stage II are $\sim$\,10$^5$ and $\sim$\,10$^6$ yr, respectively \citep{kenyon_1990,evans_2009}. This indicates that our final MYSO list is biased towards young and embedded sources, which is a natural outcome of the selection criteria of the high probable YSO candidate lists (\citealt{whitney_2008}; Sec. \ref{sec:completeness}). The quoted timescales overestimate the age of our sample since we are tracing the high-mass objects ($M$\,$\geq$\,8 M$_\odot$, Log($L$/$L_\odot$)\,$\gtrsim$\,3.5). For example, the embedded phase for MYSOs with Log($L$/$L_\odot$)\,$\textgreater$\,5.0 may only last for $\textless$\,10$^5$ yr \citep{mottram_2011}. In the remainder of this work, we will assume an age of our MYSO sample of $\sim$\,10$^5$ yr, although this number does not directly enter our analysis. 

\subsection{Completeness test}\label{sec:completeness}

Completeness of the YSO catalogues has been evaluated through false source extraction tests for both the SAGE \citep{gruendl_2009} and HERITAGE \citep{meixner_2013} data. Completeness is mainly limited by the sensitivity of the surveys and the level of the background, which predominantly hampers the detection of faint (i.e., low-mass) YSOs. Luckily, MYSOs are expected to be among the brightest sources detected in the mid-to-far IR \citep{whitney_2008, seale_2014}. 

\begin{figure}
\centering
\includegraphics[width=8.5cm]{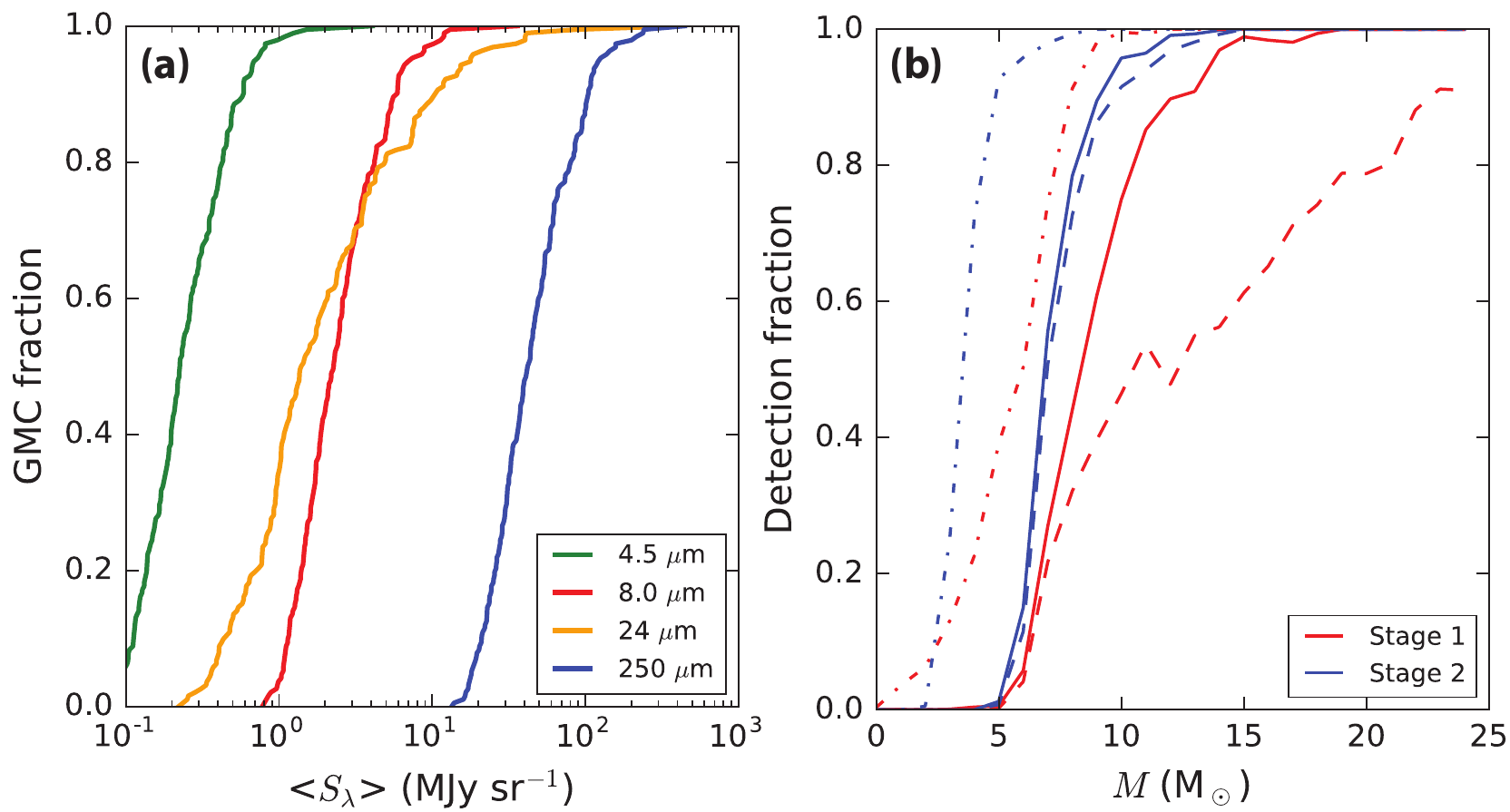} 
\caption{Surface brightness of GMCs and the detection fraction of YSO sources at the distance of the LMC. {\bf (a)}: the average surface brightness of GMCs identified in the LMC (Sec. \ref{sec:dendrogram}).  {\bf (b)}: predicted detection fraction of YSOs as a function of source mass in different of regimes of background level: `LMC average' {\em (dash-dotted lines)}, `high background' {\em (solid lines)} and the limiting case of `extreme background' {\em (dashed lines)}. See text for explanation. Results are shown for Stage 1 and 2 YSOs.}
\label{fig:detection}
\end{figure}

\citet{gruendl_2009} and \citet{meixner_2013} provide completeness limits as a function of background emission level. For SAGE, we take the completeness limits as given by \citet{gruendl_2009} for the highest backgrounds the authors were able to trace, $\sim$\,10 MJy sr$^{-1}$, as well as the LMC average, resulting in 1.8 (0.1), 3.1 (0.2), 3.5 (0.2), 5.8 (0.4), 24 (1.5) mJy for high background (LMC average) at 3.6, 4.5, 5.8, 8.0, and 24 $\mu$m, respectively. Similarly, \citet{meixner_2013} provide completeness limits in high background ($\textgreater$\,2.5 - 25 MJy sr$^{-1}$, depending on the photometric band) and the LMC average of the HERITAGE images, corresponding to 450 (450), 400 (160), 300 (60), 400 (60), 400 (100) mJy for high background (LMC average) at 100, 160, 250, 350, 500 $\mu$m, respectively. This completeness limit was shown by \citet{seale_2014} to be valid up to at least 10$^2$ MJy sr$^{-1}$ for the 250 $\mu$m band. In this work, we are particularly interested in quantifying the expected detection fraction of YSOs within GMCs. As molecular clouds may lie amid bright dust emission associated with star formation, we will investigate the detection fraction of YSOs in regions of high background. Figure \ref{fig:detection}a shows that the average surface brightness $\textless$$S_\m{\lambda}$$\textgreater$ of the vast majority of GMCs identified in the LMC ($\sim$\,200 in total; Sec. \ref{sec:dendrogram}) lie below the threshold identified for `high background' (10 MJy sr$^{-1}$ for the 4.5, 8.0, 24 $\mu$m bands, 10$^2$ MJy sr$^{-1}$ for the 250 $\mu$m band).

We use the \citet{robitaille_2006} YSO models to translate from flux to mass space: using the models, we predict the observed flux from a YSO at the distance of the LMC, which we compare with the aforementioned completeness limits at a given background level. We consider a YSO to be detected if the predicted flux from the model exceeds that of our completeness limit in {\em at least} three photometric bands. We do not include the 2MASS and MIPS 70 $\mu$m filters in the completeness test since for these bands the completeness limits have not been investigated. This also means that we cannot address the completeness for Stage 3 sources, which predominantly emit at optical to near-IR wavelengths (i.e., 2MASS). Figure \ref{fig:detection}b shows that the SAGE/HERITAGE observations should be most sensitive to Stage 2 sources, but the stringent color cuts applied by \citet{whitney_2008} and \citet{gruendl_2009} to separate YSOs from foreground and background contaminants renders our census of Stage 2 (and Stage 3) sources incomplete. However, these sources are largely irrelevant to this work since we aim to probe {\em youngest} population of YSOs, i.e., the {\em earliest} stages of star formation. Indeed, the `allowed' mid-IR color space encompasses the predicted colors of Stage 1 sources \citep{whitney_2008}: the youngest, most embedded sources that shine brightly in the mid-to-far IR, presumably as they did not have time to dissipate their surrounding material. 

We consider the detection fraction of Stage 1 MYSOs ($M$\,$\geq$\,8 M$_\m{\odot}$). Figure \ref{fig:detection}b shows that, averaged over the LMC, we recover $\sim$\,90\% of the Stage 1 MYSOs. Even within regions of high background (which again is not representable for our entire sample; Fig. \ref{fig:detection}a), we recover the majority ($\textgreater$\,50\%) of Stage 1 MYSOs, a fraction that quickly rises with source mass $M$. Finally, we consider the limiting case of 30 Doradus. At short wavelengths, diffuse background emission from warm dust and/or PAHs can arise in areas that are highly-illuminated by nearby massive stars (e.g., clouds near 30 Doradus), especially at 8 $\mu$m and 24 $\mu$m. At far-IR wavelengths, the emission from the diffuse ISM or cold dust can become significant, while the increasing beam size towards longer wavelengths will decrease the contrast of a (point-like) YSO with its surroundings. To estimate our ability to detect MYSOs in the extreme background of the 30 Doradus region, we raise the surface brightness thresholds for the `high background' regions (see above) by an order of magnitude, assume that the completeness limits follow surface brightness linearly, and re-evaluate our detection fractions. Visual inspection of the SAGE images reveal that even in the 30 Doradus region the background does not exceed $\sim\,$10 MJy sr$^{-1}$ at 3.6, 4.5, and 5.8 $\mu$m, and therefore we do not raise the completeness limits for these bands. Figure \ref{fig:detection} shows that even within this case of extreme background, we recover the majority of Stage 1 source of $M$\,$\gtrsim$\,10 M$_\m{\odot}$. 

\begin{figure*}
\centering
\includegraphics[width=18cm]{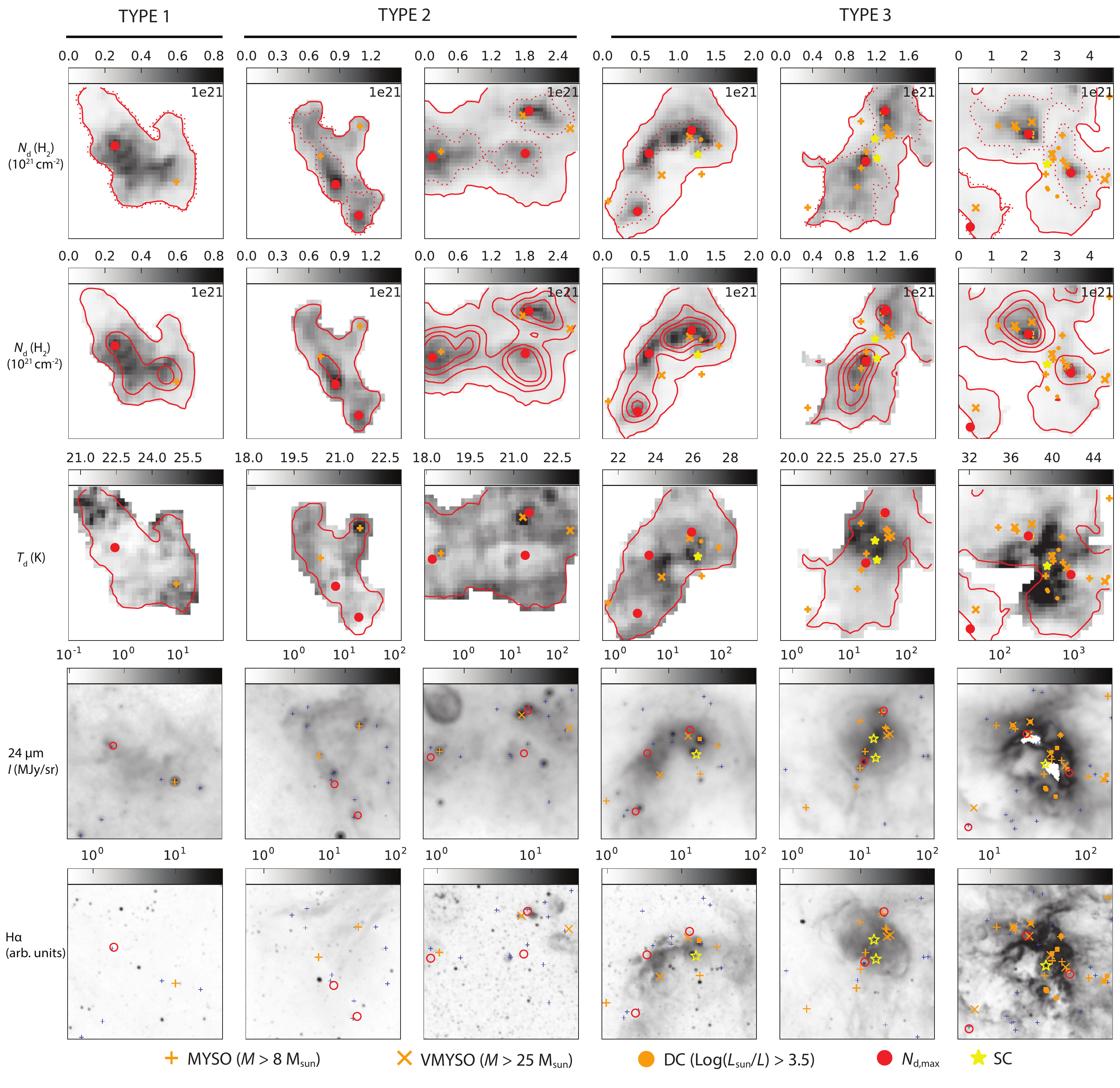} 
\caption{The internal structure of GMCs and its relation to stellar sources. Examples are shown for Type 1 (GMCs with no associated \HII\ region and/or SC; \citealt{kawamura_2009}), Type 2 (GMCs with associated \HII\ regions), and Type 3 (GMCs with associated \HII\ regions {\em and} optical stellar clusters). Overplotted in all panels: massive young stellar objects (MYSOs; {\em orange plus symbols}), {\em very} massive young stellar objects (VMYSOs; {\em orange cross symbols}), dust clumps (DCs; {\em orange dots}), and optical stellar clusters ({\em yellow asterisks}; \citealt{kawamura_2009}). {\em First row}: total column density of H$_2$ as revealed by FIR emission. We show the lowest level of the dendrogram structure (`islands'; {\em red solid line}), defined as the 3$\sigma$ sensitivity limit of the MAGMA survey \citep{wong_2011}. The highest level of substructure identified by the dendrogram algorithm within the islands are overplotted (`clouds'; {\em dotted red line}) as well as the location of its peak value (`$N_\m{max}$'; {\em red dot}).  {\em Second row:} same image as in the top row, but now overlaid with $^{12}$CO(1-0) emission contours, offering an independent measure of column density. The gas and dust are found to be in good agreement with each other. {\em Third row:} dust temperature maps, overlaid with island footprints and locations of $N_\m{max}$. {\em Fourth row}: MIPS 24 $\mu$m intensity maps (see the online version for a high resolution of these figures). In this panel, we have removed the filling of the red dots ($N_\m{max}$) and yellow asterisks (SCs) to better reveal their background, and we have added sources from the entire catalogue of YSOs {\em (blue plus symbols)}, which contain sources either below our completeness limit or those that remain uncharacterized (Sec. \ref{sec:catalogue}). Depending on their evolutionary state, MYSOs are seen in sharp contrast with their surroundings at 24 $\mu$m (Sec. \ref{sec:completeness}), except in a region of extreme background, such as the 30 Doradus region (outermost right panel), part of which is saturated in the 24 $\mu$m band. {\em  Fifth row}: MCELS \citep{smith_1998} H$\alpha$ images (uncalibrated), revealing that Stage 1 MYSO do not have an optical counterpart, confirming the embedded nature of these sources. All image span $\sim$\,100\,$\times$\,100 pc at the distance of the LMC.}
\label{fig:postage}
\end{figure*}

The success of recovering MYSOs can be attributed to the large contrast of MYSOs with the ISM at mid-IR wavelengths (see Fig. \ref{fig:postage}). Note that many of the LMC MYSOs may eventually break up in small clusters given our limited resolution, however it is expected that the source luminosity is dominated by its highest-mass member since $L$\,$\propto$\,$M^\m{\alpha}$, with $\alpha$ $\textgreater$ 1 \citep{tout_1996}. We conclude that our catalogue of YSOs should be complete for Stage 1 sources of $M$\,$\geq$\,8 M$_\m{\odot}$ as they are bright enough to be detected in the SAGE/HERITAGE surveys, although some sources without a point source counterpart in the mid-IR may have eluded detection within {\em extreme} regions of IR background. The analysis in the remainder of this work is based exclusively on the 569 Stage 1 MYSOs.

\section{Giant molecular clouds: column densities \& substructure}\label{sec:mapandcloud}

In this work, we are particularly interested in characterizing the distribution of material within GMCs and its relation to the MYSOs. Column density maps can be derived either from the far-IR HERITAGE images (dust-based) or the $^{12}$CO(1-0) emission from the MAGMA survey (gas-based). 

It is well known that $^{12}$CO(1-0) emission alone is an unreliable tracer of mass {\em concentration} within GMCs harboring massive star formation, since there are numerous pathways that may affect the $^{12}$CO(1-0) emission. The principle advantage of dust over gas in column density estimates is the dynamic range probed; gas tracers are only sensitive to a specific range in volume densities that relate to critical densities, depletion, and opacity effects. In addition, \citet{madden_1997} found evidence for hidden molecular hydrogen not traced by CO in low-metallicity irregular galaxies using the [\CII\,] 158 $\mu$m line, while \citet{bernard_2008} noted a hidden molecular phase (i.e., `CO-dark gas') through the presence of a FIR excess emission that can not be explained through \HI, and yet does not correlate with CO emission. This hidden phase may be significant in the low-metallicity environments of the Magellanic clouds \citep{jameson_2015}. Lastly, heating by young massive stars may affect the CO emissivity per unit mass \citep{scoville_1987}, questioning the validity of a `constant' $X_\m{CO}$ factor within regions of massive star formation. We conclude that for the purpose of this study, using the FIR dust emission to trace molecular gas represents a more direct and robust method that avoids the known biases of CO as a tracer of H$_2$ and, therefore, we proceed our investigation by solely using dust-based column density maps.

\subsection{Far-infrared column densities}\label{sec:column}

We fit the dust far-IR SED on a pixel-to-pixel basis (pixel size 10") assuming optically thin emission using a single-temperature blackbody modified by a power law emissivity, $I_\m{\lambda}$ = $\Sigma_\m{d}$$\kappa$($\lambda$,$\beta$)$B_\m{\lambda}$($T_\m{d})$. Here, $\Sigma_\m{d}$ is the dust surface density, $B_\lambda$ is the Planck function, $\kappa_\m{\lambda}$ = ($\kappa_\m{eff}$/160$^{-\beta}$)$\lambda^{-\beta}$ the emissivity law with $\kappa_\m{\lambda}$ the dust emissivity at wavelength $\lambda$, $\kappa_\m{eff}$ = 28.9 cm$^2$ g$^{-1}$ the dust emissivity at reference wavelength $\lambda$ = 160 $\mu$m \citep{gordon_2014}, and $\beta$ the spectral index. Note that the value for $\kappa_\m{eff}$ is larger than the value given in \citet{gordon_2014}; the reported value was erroneously tabulated as $\kappa_\m{eff}$/$\pi$. This error did not propagrate into the analysis or results (K. D. Gordon, priv. communication). We fit the HERITAGE photometric data following the method described in \citet{gordon_2014}, and leave $\Sigma_\m{d}$, $T_\m{d}$, and $\beta$ as free parameters. The \citet{gordon_2014} fit method reduces the degeneracy between, e.g., $T_\m{d}$ and $\beta$ \citep{dupac_2003,shetty_2009} by accounting for the correlated errors between the {\em Herschel} bands, while using the {\em full} likelihood function for each parameter (i.e., the expectation value), as opposed to $\chi^2$ minimizations that only use the maximum value of the likelihood. 

The submilimeter excess, defined as the excess emission seen at submillimeter wavelengths above that expected for dust grains at a single temperature and  $\lambda^{-\beta}$ emissivity law (i.e., our adopted model), contributes 27\% to the observed 500 $\mu$m flux averaged over the entire LMC \citep{gordon_2014}. Based on observed gas-to-dust ratios, \citet{gordon_2014} argue that the submillimeter excess is more likely to be due to emissivity variations than a second population of very cold dust. For this reason, we have opted to exclude the SPIRE 500 $\mu$m band in our model fitting: while we sacrifice a data point on the Rayleigh-Jeans part of the SEDs, we avoid contamination by submillimeter excess emission that can not be captured by our single-temperature model. At the same time, this choice increases the resolution of our dust model maps from $\sim$\,36" (9 pc) to $\sim$\,25" (6 pc), which constitutes a significant improvement to the cause of our study, since we aim to relate the location of MYSOs to the internal structure of GMCs. From $\Sigma_\m{d}$, we convert to molecular hydrogen column density through $N$(H$_2$) = $R$$\Sigma_\m{d}$/$\mu_\m{H}$$m_\m{H}$, where $R$ is the gas-to-dust ratio in the LMC ($\approx$ 380; \citealt{roman-duval_2014}), $m_\m{H}$ is the mass of a hydrogen atom, and we take $\mu_\m{H}$ = 2.8 as mean molecular weight per hydrogen molecule. 

\subsection{Deconvolution of GMCs}\label{sec:dendrogram}

We chose to deconvolve the hierarchical structure of the dust column density of GMCs using the dendrogram technique \citep[see][]{rosolowsky_2008}. Dendrograms trace local significant maxima and the way these maxima are connected along isocontours. Compared to other cloud-decomposing algorithms such as CLUMPFIND \citep{williams_1994}, dendrograms have been shown to be more robust against noise and user-defined parameters \citep{goodman_2009,pineda_2009}. The HERITAGE data suffer from residual striping effects along the PACS/SPIRE scan directions that propagate as fluctuations in our column density maps with a level of $\Delta$$N$(H$_2$)\,$\sim$\,1\,-\,4\,$\times$\,10$^{20}$ cm$^{-2}$. To avoid being biased by residual artifacts in our column density maps, we define local maxima ($N_\m{max}$) as a structure that has a minimum column density contrast of $\Delta$$N$({H$_2$}) = 8\,$\times$\,10$^{20}$ cm$^{-2}$, while containing a minimum number of pixels that exceeds the beam area of the HERITAGE survey (Sec. \ref{sec:observations}) by a factor of 2. Since we are interested in the properties of GMCs, we limit the deconvolution of the dust-based column density maps to regions of the LMC that exhibit significant CO emission, which we obtain from MAGMA integrated intensity maps with a rms noise level of $\sigma_\m{noise}$ $\sim$\,0.4 K km s$^{-1}$ \citep{wong_2011}. Note that by restricting the dust maps to MAGMA positive detections we may exclude the more diffuse areas of GMCs projected on the sky, i.e., the `CO-dark' phase (Sec. \ref{sec:mapandcloud}). However, in this work we are tracing the formation of massive stars within GMCs, which is very likely to occur in high volume density regimes of GMCs at column depths large enough for CO to survive (`CO-bright' regions) 

We follow the nomenclature of \citet{wong_2011} and \citet{hughes_2013} and refer to the largest contiguous structures of CO emission detected by MAGMA as `islands'. The internal column densities of individual islands are subsequently derived using the higher-resolution dust-based column density maps (see Fig. \ref{fig:postage} for several examples). Note that at higher resolution, the LMC GMCs appear less extended than implied by the CO-based islands, since $N$(H$_2$)\,$\lesssim$\,10$^{21}$ cm$^{-2}$ in parts of the island footprints, which are regions where CO is expected to be dissociated and not detectable (e.g., \citealt{bolatto_2013}). Moreover, the dust-based column density maps show large-scale density enhancements, from localized density peaks to filaments of tens of pc in size. We define a `cloud' as the highest column density structure within an island as identified by the dendrogram analysis (Fig. \ref{fig:postage}). In the remainder of this paper, `islands' and `clouds' exclusively refer to the products of our dendrogram decomposition.

\section{The distribution of massive star formation within giant molecular clouds}\label{sec:results}

\citet{kawamura_2009} classified GMCs in the LMC as Type 1 (GMCs with no massive star formation), Type 2 (GMCs with associated \HII\ regions), and Type 3 (GMCs with associated \HII\ regions {\em and} optical stellar clusters). This classification was based on GMCs detected in the NANTEN survey (at resolution 2.6'; \citealt{fukui_2008}), not all of which have been observed by MAGMA (at resolution 45"; \citealt{wong_2011}), but a lot of which reveal substructure at higher resolution. To directly compare with the work of \citet{kawamura_2009}, we consider all MAGMA islands detected within the footprint of a NANTEN GMC as being of the same `Type'. The results are shown in Tab. \ref{tab:mysos}. We find 42, 93, and 52 Type 1 islands, Type 2 islands, and Type 3 islands, which are further decomposed into 72, 160, 213 individual clouds. We note that 74 individual structures are not matched. These unmatched structures partially represent structures observed with MAGMA that were not detected with the NANTEN survey, but are mostly small fragments that fall outside of the ellipsoid footprints defined by the NANTEN catalogue \citep{fukui_2008} and are separated from an island at the higher resolution of MAGMA. Indeed, even though appearing large in number, the combined CO mass incorporated in these 74 fragments is only 4\% of the entire luminous CO mass detected by MAGMA. Given this, we exclude these unmatched structures to our further analysis, as we expect they will not affect the conclusions in this paper. 

We recover a total of 311 MYSOs within the CO island boundaries (out of 569 total; Sec. \ref{sec:catalogue}), which implies that almost 50\% of MYSOs have not been associated with CO emission in the MAGMA survey, which was already noted by \citet{wong_2011}. This mainly results from the incomplete coverage of the MAGMA survey, but also because the survey is insensitive to clouds below $M$\,$\sim$\,3\,$\times$\,10$^4$ M$_\m{\odot}$, as well as possible molecular cloud disruption through massive star feedback (see for a discussion \citealt{wong_2011}). 
 
The average number of MYSOs per island, $\textless$$N_\m{MYSO}$$\textgreater$, equals 0.2, 1.0, and 3.9 for Type 1, Type 2, and Type 3 islands, respectively (Tab. \ref{tab:mysos}). The percentage of Type 1 islands that shows at least one MYSO is 14\% (an example of which is shown in Fig. \ref{fig:postage}). This means that the classification by \citet{kawamura_2009} is largely consistent with our MYSO census, a result that may be surprising given that the authors did not include in their analysis the young, dust enshrouded phase of star formation revealed by the SAGE and HERITAGE surveys. The quantities $p_\m{MYSO}$ and $\textless$$N_\m{MYSO}$$\textgreater$ increases in Type 2 and Type 3 islands, confirming that these regions are more actively forming massive stars \citep{kawamura_2009}. Note that these numbers shift downwards when considering clouds as opposed to islands, which implies that massive star formation occurs in specific parts of islands, rendering the majority of clouds devoid of any (Stage 1) MYSO. Indeed, averaged over the entire galaxy, only 33\% of the LMC clouds (48\% when considering islands) show evidence for {\em ongoing} massive star formation over the past $\sim$\,10$^5$ yr (estimated lifetime of Stage 1 MYSOs; Sec. \ref{sec:catalogue}).

\begin{table}
\centering
\caption{Embedded massive star formation in GMCs}
\begin{tabular}{l|c|c|c}\hline 
 & Number & $p_\m{MYSO}$ & $\textless$$N_\m{MYSO}$$\textgreater$ \\ \hline
Type 1 (island) & 42 & 14\% & 0.2 \\ 
Type 2 (island) & 93 & 49\% & 1.0 \\  
Type 3 (island) & 52 & 75\% & 3.9 \\ \hline
Type 1 (cloud) & 72 & 9\% & 0.1 \\ 
Type 2 (cloud) & 160 & 33\% & 0.6 \\  
Type 3 (cloud) & 213 & 42\% & 0.9 \\ \hline
\end{tabular}
\tablecomments{Listed are parameters for `islands' and `clouds': total number of islands/clouds found in the dendrogram-based decomposition (Sec. \ref{sec:dendrogram}), the percentage of islands/clouds with an embedded MYSO, $p_\m{MYSO}$;  the mean amount of MYSOs per island/cloud, $\textless$$N_\m{MYSO}$$\textgreater$. The different GMC `Types' stem from the classification of \citet{kawamura_2009}.}
\label{tab:mysos}
\end{table}

\subsection{Location of massive YSOs in GMCs}\label{sec:distribution}

Figure \ref{fig:postage} compares GMC column density maps with our YSO catalogue. We plot the locations of stellar sources for several subgroups in which we have estimated that our census is complete (Sec. \ref{sec:completeness}). The first group are the MYSOs of $M$\,$\geq$\,8 M$_\m{\odot}$ (main sequence mass of $\sim$\,B2V star; \citealt{mottram_2011}). From the MYSOs, we take the most luminous sources and define a subset of {\em very} massive young stellar objects (`VMYSOs') of $M$\,$\geq$\,25 M$_\m{\odot}$ (main sequence mass of $\sim$\,O7.5V star). Note that the division of MYSOs and VMYSOs may be equivalent to separating the progenitors of B and O stars, respectively. Lastly, we include the `DCs'; given that we have not been able to derive masses for this class (Sec. \ref{sec:catalogue}), we rely on the luminosity of the source and translate this to a main sequence mass. We include DCs with Log($L$/$L_\m{\odot}$)\,$\geq$\,3.5, the luminosity equivalent of a $M$\,$\geq$\,8 M$_\m{\odot}$ main sequence star \citep{mottram_2011}. Together, these subsets define a complete tally of the youngest ($\sim$\,10$^5$ yr) sources on their way of becoming massive stars.

Figure \ref{fig:postage} shows that the positions of MYSOs, VMYSOs, and DCs do not seem to correlate well with the local column density peaks $N_\m{max}$. Upon close inspection, one may argue that MYSOs tend to {\em avoid} the highest column densities within GMCs, and are instead often positioned against the outskirts or edges of column density enhancements within clouds/islands. These density enhancements have typical sizes up to tens of pc, similar to infrared dark clouds in the Galaxy (IRDCs; \citealt{rathborne_2006}). To quantify the relative distribution between MYSOs and $N_\m{max}$, we cross-match the locations of our catalogue (Sec. \ref{sec:catalogue}) with those of $N_\m{max}$ (Sec. \ref{sec:mapandcloud}). For each MYSO, we find its nearest on-sky $N_\m{max}$, after which we plot the column density ratio at the location of the YSO, $N_\m{YSO}$, over that of its matched column density peak, i.e., $N_\m{YSO}$/$N_\m{max}$. A value of 0.0 of this ratio would mean that the source is located {\em outside} of an island (we only count MYSOs located within an island), whereas a value of 1.0 means that the source is located within the pixel containing $N_\m{max}$. 

\begin{figure*}[!t]
\centering
\includegraphics[width=18cm]{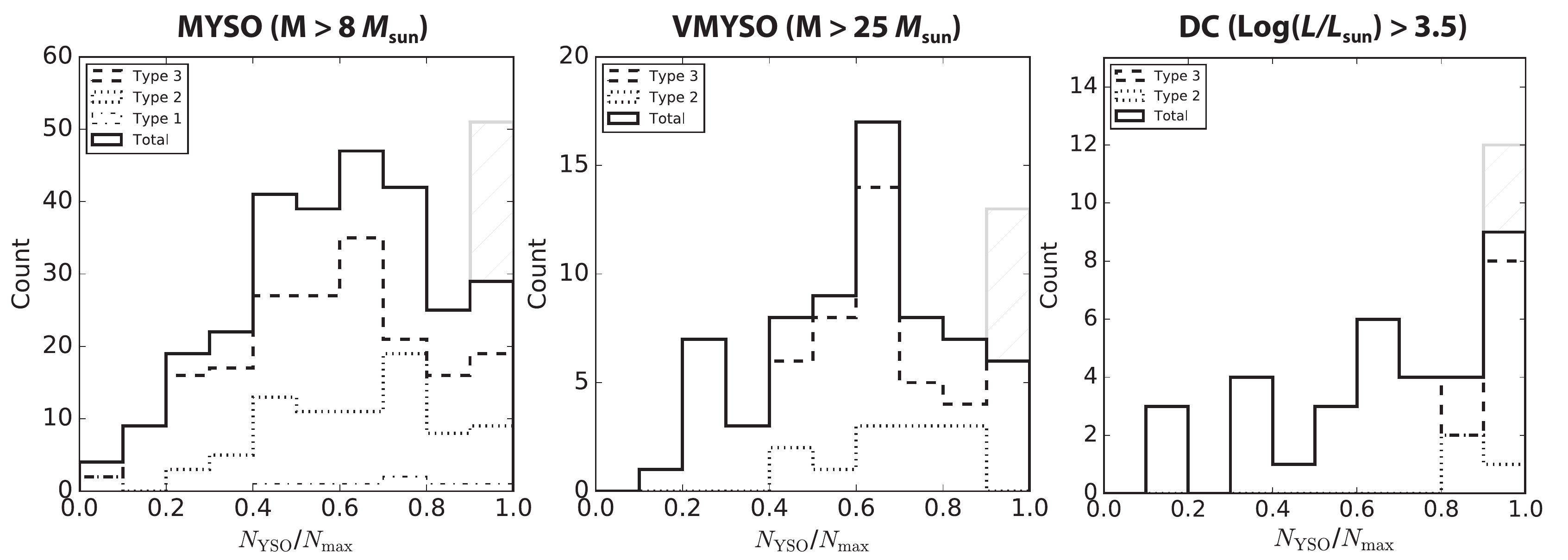} 
\caption{Relative column densities of MYSOs {\em (left panel)}, VMYSOs {\em (middle panel)}, and DCs {\em (right panel)} with respect to local H$_\m{2}$ column density peaks traced by FIR dust emission ($N_\m{max}$; Fig. \ref{fig:postage}). The results are shown for Type 1, Type 2, and Type 3 clouds, as well as the total distribution. The grey hatched area are sources that fall within the pixel of $N_\m{max}$, and are thus unresolved in this positional analysis (see text).}
\label{fig:histogram}
\end{figure*}

Figure \ref{fig:histogram} shows the results of this analysis, where we further break up the results in Type 1, Type 2, and Type 3 regions. The trend seen in Fig. \ref{fig:postage} is immediately revealed: there is a clear deficit of sources at high column densities. This holds for both the MYSO as the VMYSO distribution, and is robust against the user-defined inputs of the dendrogram decomposition (Sec. \ref{sec:mapandcloud}). The DCs are hampered by small number statistics, but this subset does seem to favor higher column densities compared to the MYSO/VMYSO distributions. The grey area in the highest column density bin are sources that fall within the pixel of $N_\m{max}$, and are thus unresolved in this positional analysis. This resolution effect stems from the relative coarse pixel scale of our column density maps (10" or 2.5 pc at the distance of the LMC), causing all sources found within the pixel of peak column density to collapse in this histogram bin. We expect that, in reality, these sources will form an extended wing of the distribution, possibly declining towards $N_\m{YSO}$/$N_\m{max}$ = 1.0. The positions of the MYSOs are known at a higher accuracy compared to our column density maps, since their detections are matched to shorter wavelength measurements including Spitzer and 2MASS \citep{meixner_2013,seale_2014}. The sources found within the pixel of $N_\m{max}$ show a smooth distribution with distance as measured from the center of $N_\m{max}$ (see below; Fig. \ref{fig:cumulative}), revealing that there is structure that is unresolved in our $N_\m{YSO}$/$N_\m{max}$ histograms.

As a caveat, we note that the exact shape of the histograms in Fig. \ref{fig:histogram} are biased by projection effects (i.e., the three-dimensional distribution of sources with respect to the GMCs), and an `aperture' effect (i.e., the effective area each bin in Fig. \ref{fig:histogram} traces). Whereas the former would only increase the dearth of YSOs towards high column densities if part of the YSOs are found at the position of $N_\m{max}$ due to change alignment along the line of sight, the latter depends on the internal density distribution of each individual cloud. A sharply peaked density profile of GMCs will cause the higher density bins of Fig. \ref{fig:histogram} to trace only a small part of the cloud in spatial terms. Still in this case, {\em if} massive stars would form at the highest column densities of GMCs, we would expect to see a strongly peaked profile skewed towards $N_\m{YSO}$/$N_\m{max}$\,$\sim$\,1.0. 

The sample size in Type 1 islands (6 in total for the MYSOs) are too small for a statistical analysis, which would have provided insight into the formation of massive stars in presumably the least evolved or youngest GMCs \citep{kawamura_2009}. For Type 2 and Type 3 clouds, we see that the distributions peak at roughly $N_\m{YSO}$/$N_\m{max}$\,$\sim$\,0.6 - 0.65. Interestingly, in Type 3 islands, the chance of finding a VMYSO in the outskirts of a cloud/island ($N_\m{YSO}$/$N_\m{max}$\,$\sim$\,0.2) is similar to that near its peak column density ($N_\m{YSO}$/$N_\m{max}$\,$\sim$\,1.0). Even though small in number, we note that DCs appear to favor high column densities.

\begin{figure}
\centering
\includegraphics[width=7.5cm]{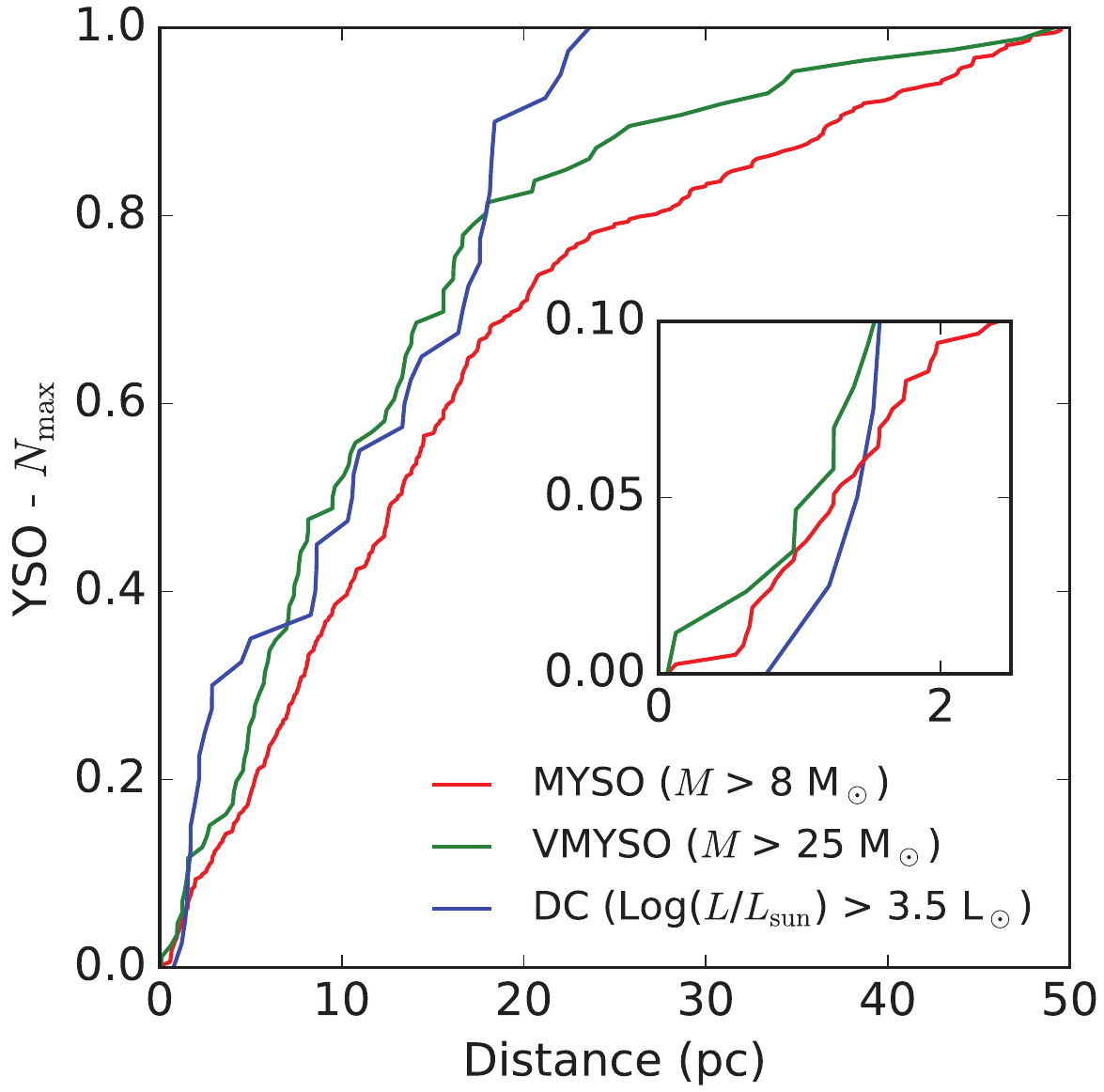} 
\caption{Cumulative distribution of the distance between YSOs and the nearest on-sky column density peaks ($N_\m{max}$). Plotted are MYSOs {\em (red)}, a subset of VMYSOs {\em (green)}, and DCs (Sec. \ref{sec:catalogue}; {\em blue}).}
\label{fig:cumulative}
\end{figure}

Figure \ref{fig:cumulative} shows the cumulative distribution of the distance between MYSOs, VMYSOs, and DCs with the nearest column density peak $N_\m{max}$. The figure shows that VMYSOs and DCs objects tend to be located closer to $N_\m{max}$ compared to MYSOs. The similarity of the distribution may indicate that DCs represent part of the VMYSO distribution that have been misclassified in our YSO catalogue (Sec. \ref{sec:catalogue}). Nonetheless, half the objects within all subsamples are found $\sim$\,10 pc away from $N_\m{max}$. The inset shows the distribution of sources within the pixel of $N_\m{max}$ (as measured from the pixel center), showing structure that is unresolved in our column density maps.

\begin{figure*}
\centering
\includegraphics[width=17cm]{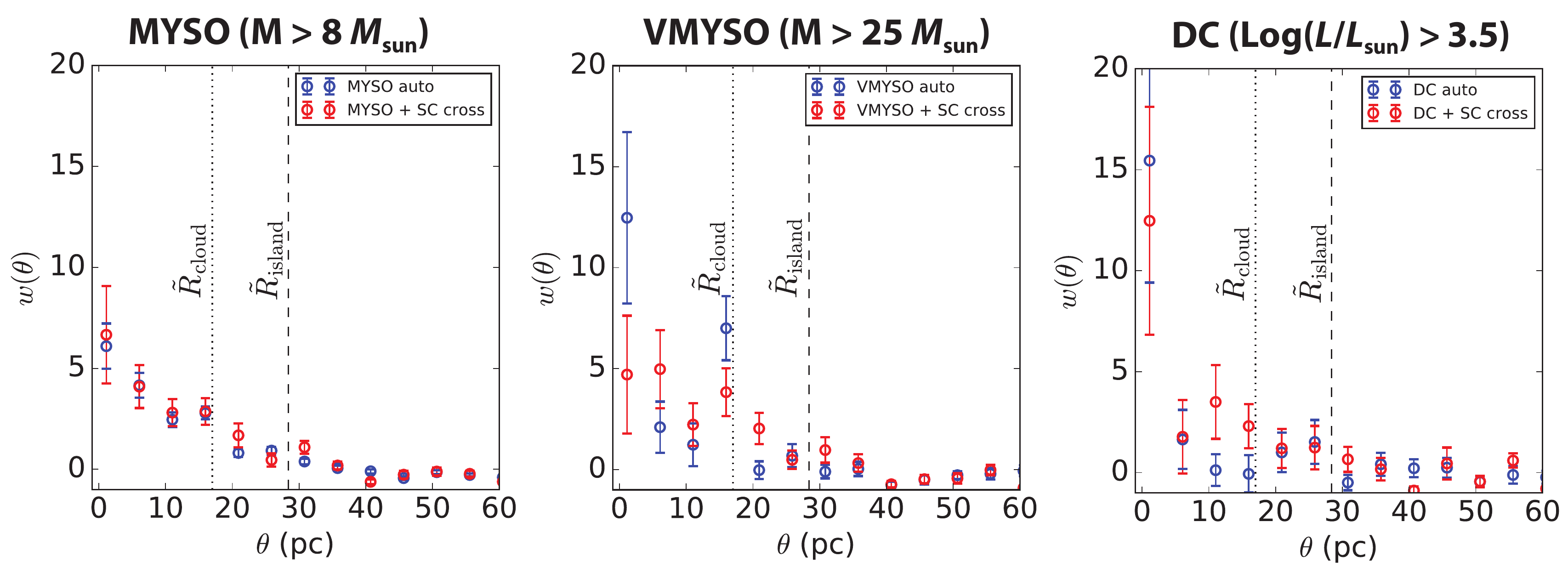} 
\caption{The clustering of massive star formation. Shown is the angular correlation function $w(\theta)$ as a function of separation $\theta$ for the auto-correlation (Eq. \ref{eq:auto}; {\em blue points}) and cross-correlation with SCs (Eq. \ref{eq:cross}; {\em red points}). The three panels show the results for MYSOs, VMYSOs, and DCs, respectively (see text).}
\label{fig:clustering}
\end{figure*}

\subsection{Clustering and the connection between different generations of massive star formation}\label{sec:propagation}

Given that we have obtained a unique set of MYSOs and GMCs throughout an {\em entire} galaxy, we are able to probe the massive star formation process as a function of environment and evolutionary stage. Moreover, the sequential behavior of massive star formation can be probed by combining our catalogue with the results of \citet{kawamura_2009}, who matched NANTEN GMCs with optical stellar clusters (SCs) younger than 10 Myr (taken from \citealt{bica_1996}) in order to define Type 3 GMCs. In the following analysis, we use positions as reported by \citet{bica_1996}, but caution the reader that the exact central positions for these clusters may not be well known since some of the sources could be extended OB associations. Nonetheless, these young SCs likely represent a more evolved generation of massive stars compared to the YSOs traced by SAGE and HERITAGE, as they have already emerged from their parent clouds to shine bright at optical wavelengths. 

\subsubsection{Angular correlation function}

We investigate the clustering of massive star formation by using an angular correlation function. The Landy-Szalay estimator \citep{landy_1993} calculates the correlation $w\m{(\theta)}$ through

\begin{equation}	
\label{eq:auto}
w(\theta) = \frac{n_\m{DD}(\theta) - 2n_\m{DR}(\theta) + n_\m{RR}(\theta)}{n_\m{RR}(\theta)}, 
\end{equation}

where $n$ represents the number of pair counts between `true' data (subscript D) and `random' data (subscript R) as a function of angular distance $\theta$. Equation \ref{eq:auto} computes the intrinsic correlation (or rather: clustering) of a dataset, i.e., its `auto-correlation', but it can be generalized for two different datasets \citep{bradshaw_2011} to compute a `cross-correlation'

\begin{equation}	
\label{eq:cross}
w(\theta) = \frac{n_\m{D_1D_2}(\theta) - 2n_\m{D_1R_2}(\theta) - n_\m{R_1D_2}(\theta) + n_\m{R_1R_2}(\theta)}{n_\m{R_1R_2}(\theta)}.
\end{equation}

\vspace{3mm}
Fundamentally, $w\m{(\theta)}$ gives the clustering of a set of points containing positional information {\em in excess} over what is expected from a random distribution of points in the same field. \citet{thompson_2012} and \citet{kendrew_2012,kendrew_2016} applied Eqs. \ref{eq:auto} and \ref{eq:cross} to the distribution of clumps and bubbles drawn from large surveys in the Galactic plane, and demonstrated the use of angular correlation functions to the study of (massive) star formation and its relation to larger scale structures in the ISM. Here, we apply a similar methodology to the MYSO (691 sources), VMYSO (101 sources), and DC (36 sources) catalogues. We used the public code by S. Kendrew \citep{kendrew_code_2015}, which makes use of the Astropy package \citep{robitaille_2013}, and adapted this code for our specific analysis. Random catalogues were constructed over the extent of the LMC (71\,$\leq$\,RA\,$\leq$\,89, -71\,$\leq$\,Dec\,$\leq$\,-65) with a sample size that is 50 times as large as the input (`true') data catalogue to ensure an adequate sampling of the covered area. The uncertainty in $w$($\theta$) is determined using 100 bootstrap resamples \citep{ling_1986}, where pseudo datasets were generated by sampling points with replacement from the (`true') input catalogue, while maintaining the same number of sources as the input catalogue. The correlation function is then calculated for each of the bootstrap samples: we report the mean and its 1$\sigma$ uncertainty. For the cross-correlations, the pair counts were normalized to account for different catalog sizes. We bin the results in steps of $\Delta$$\theta$ = 5 pc, and start our binning at 1 pc to exclude pair counts of data with themselves. In each panel, we show the auto-correlation of the respective samples, as well as their cross-correlation with the SC sample.

Fig. \ref{fig:clustering} shows the results of this routine. We compare the correlations to the median radius of all islands ($\tilde{R}_\m{island}$ = 28.4 pc) and clouds ($\tilde{R}_\m{cloud}$ = 17.0 pc) identified by the dendrogram algorithm (Sec. \ref{sec:mapandcloud}), where the radii are estimated assuming spherical symmetry through $R$ = $\sqrt{A/\pi}$, where $A$ represents the surface area of the island/cloud in pc$^2$. Not surprisingly, all correlations show clustering ($w$\,$\textgreater$\,0) at $\theta$\,$\leq$\,$\tilde{R}_\m{island}$ and $\theta$\,$\leq$\,$\tilde{R}_\m{cloud}$, implying that massive star formation predominantly occurs within the boundaries of an island or cloud. The error bars for the VMYSOs and DCs are relatively large because of the small sample size. Note, however, that the strongest clustering for all sources occurs towards the smallest scales, i.e., $\theta$\,$\leq$\,10 pc, significantly less than $\tilde{R}_\m{island}$ and $\tilde{R}_\m{cloud}$. The cross-correlations with SCs show similar trends compared to the respective auto-correlations, suggesting that MYSOs, VMYSOs, and DCs all reside close to SCs. In summary, we conclude that massive star formation clusters on scales much smaller than the size of parent islands and/or clouds, and that this clustering holds over different generations on timescales up to 10 Myr \citep{bica_1996}.

\begin{figure}
\centering
\includegraphics[width=8.5cm]{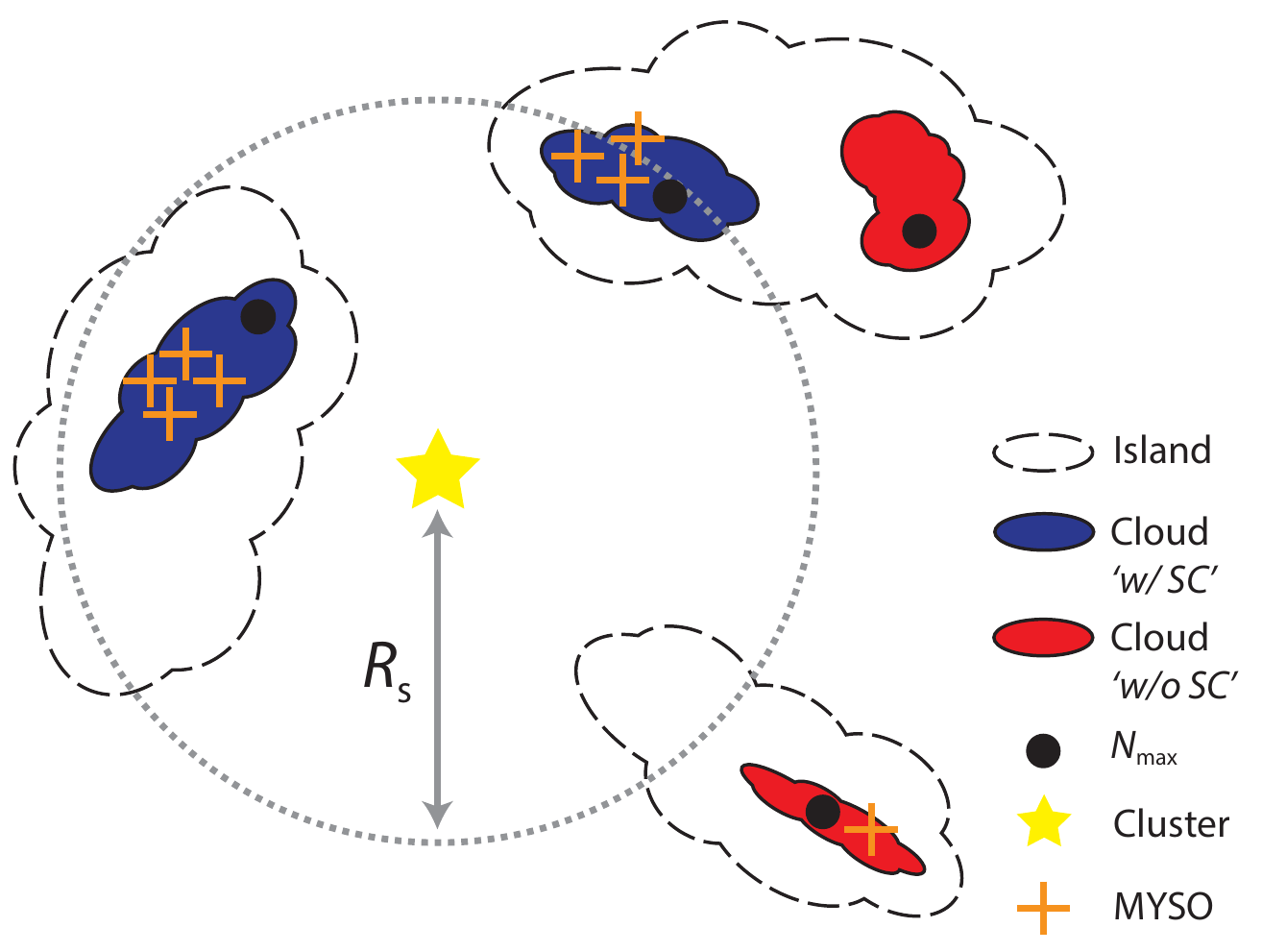} 
\caption{Cartoon depicting the method associating stellar clusters (SCs) with clouds. Our cloud decomposition (Sec. \ref{sec:dendrogram}) distinguishes between `islands' {\em (long dashed lines)}, `clouds' {\em (solid lines)}, and column density peaks $N_\m{max}$; see also Fig. \ref{fig:postage}. We define a search radius, $R_\m{s}$, around each SC of our sample. We then separate $N_\m{max}$ and their parent clouds that fall within this search radius (`w/ SC'; {\em blue clouds}), from those that fall outside of the search radius (`w/o SC'; {\em red clouds}). We then compare the amount of MYSOs/VMYSOs/DCs found in both samples (Fig. \ref{fig:triggering}).}
\label{fig:cartoon}
\end{figure}

\begin{figure*}
\centering
\includegraphics[width=18cm]{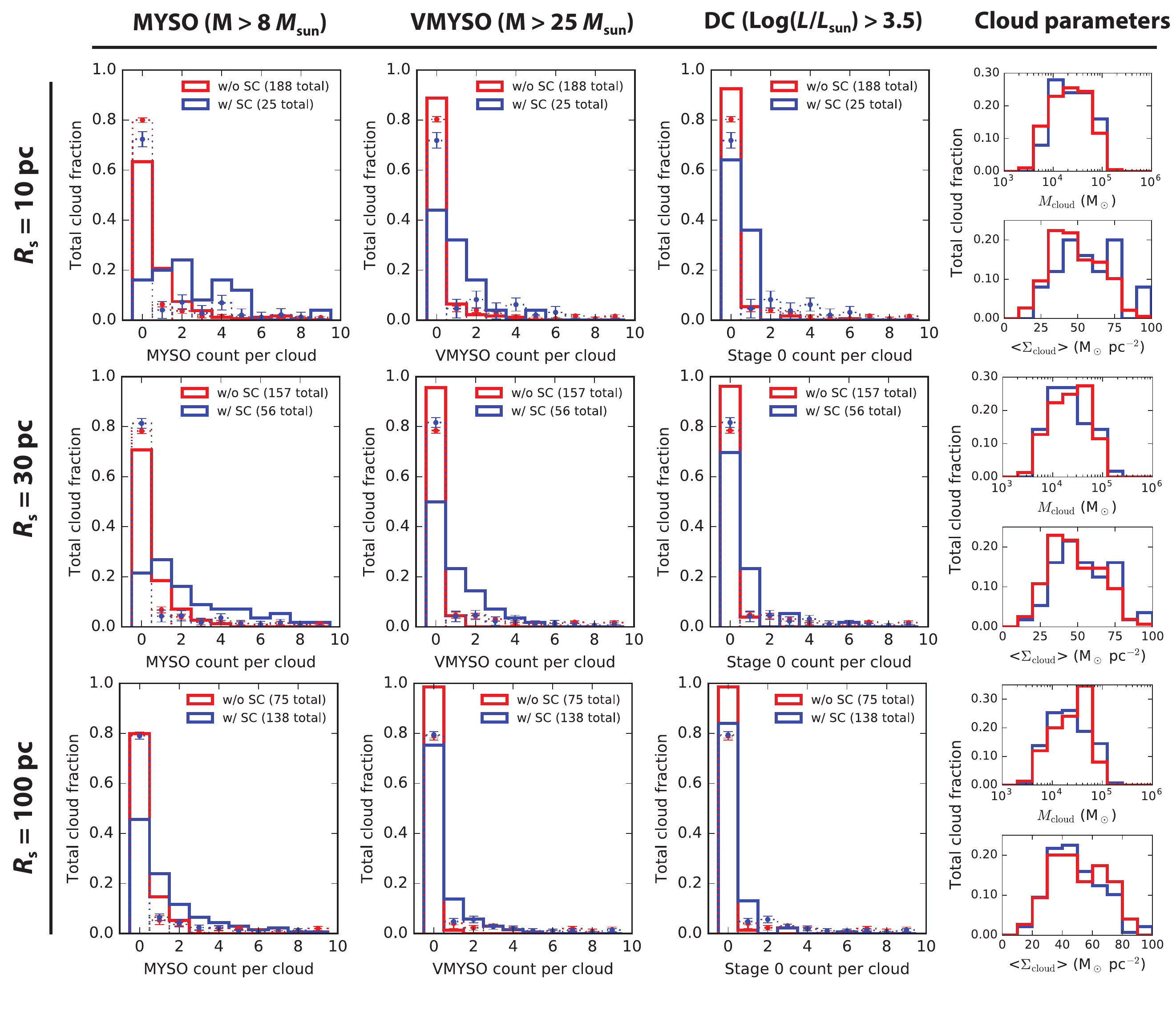} 
\caption{The presence of stellar clusters and the rate of massive star formation in Type 3 clouds. Plotted are results for MYSOs, VMYSOs, and DCs for different search radii $R_\m{s}$: 10, 30, 100 pc (see text and Fig. \ref{fig:cartoon}). We separate $N_\m{max}$ (and their associated parent clouds/islands; Fig. \ref{fig:postage}) at distances d $\leq$ $R_\m{s}$ (`w/ SC'; {\em (blue solid histogram}) from those at d $\geq$ $R_\m{s}$ (`w/o SC'; {\em red solid histogram}), and count the amount of objects (MYSOs, VMYSOs, DCs, respectively) located within each individual cloud. Histograms are normalized to their respective total number count of clouds (shown in upper right) to compute the total cloud fraction. The results are compared to a situation where the same number of objects (MYSOs, VMYSOs, DCs) are distributed randomly within islands: the dotted histograms show the mean and 1$\sigma$ uncertainty of 100 randomizations. The far right column show the distributions of mass $M_\m{cloud}$ and average surface density $\textless$\,$\Sigma_\m{cloud}$\,$\textgreater$ of the parent clouds.}
\label{fig:triggering}
\end{figure*}

\subsubsection{The presence of stellar clusters and the rate of massive star formation in clouds}

An alternative way of quantifying the connection between multiple generations of massive star formation is illustrated in Fig. \ref{fig:cartoon}. We define a search radius $R_\m{s}$ around each SC, along with a distance $d$ between a SC and each peak column density $N_\m{max}$. Stellar clusters are then matched with $N_\m{max}$ (and their parent cloud) if they fall within the defined value of the search radius, i.e., $d$\,$\leq$\,$R_\m{s}$. This routine provides us with a set of $N_\m{max}$ and parent clouds that are located within $R_\m{s}$ of SCs (clouds with SCs; `w/ SC'), and a set that falls outside of the search radius around SCs (clouds without SCs; `w/o SC'). After this, we count the amount of MYSOs/VMYSOS/DCs associated with each individual cloud. The advantage of this analysis over the use of the angular correlation functions (Eqs. \ref{eq:auto} and \ref{eq:cross}) is that we can quantify the connection between MYSOs/VMYSOs/DCs and SCs, while directly relating this to the parent molecular cloud and its global properties, such as mass ($M_\m{cloud}$) and average surface density ($\mean{\Sigma_\m{cloud}}$). To test the significance of our findings, we compare the results with an identical analysis, but using data where MYSO/VMYSO/DCs have been distributed randomly within islands, keeping the total number of sources per island constant.

In the following analysis, we will use search radii of $R_\m{s}$ = 10, 30, and 100 pc. Note that by using the SC catalogue from \citet{kawamura_2009}, we limit the analysis to Type 3 GMCs. Indeed, we find a median $\bar{d}$ $\sim$\,370 pc, 280 pc, and 60 pc for Type 1, 2, and 3 clouds, respectively, confirming the close association of these SCs with Type 3 GMCs. 

\begin{table}
\centering
\caption{}
\begin{tabular}{l|c|c|c}\hline 
 & $R_\m{s}$ = 10 pc & $R_\m{s}$ = 30 pc  &$R_\m{s}$ = 100 pc \\ \hline
MYSOs & 31\% (13\%) & 65\% (18\%)  & 91\% (51\%) \\  
VMYSOs & 39\% (8\%) & 89\% (18\%) & 98\% (52\%) \\  
DCs & 26\% (9\%) & 82\% (14\%) & 97\% (47\%) \\ \hline
\end{tabular}
\tablecomments{Percentage of MYSOs/VMYSOs/DCs associated with clouds found within search radius $R_\m{s}$. In parentheses we provide the same numbers derived through 100 randomizations of the same number of objects (MYSOs, VMYSOs, DCs) within islands}.
\label{tab:yield}
\end{table}

The results are shown in Fig. \ref{fig:triggering}. We note upfront that if the presence of SCs would increase the amount of MYSOs/VMYSOs/DCs in clouds, we would expect to see a larger object count per cloud in the `w/ SC' sample compared to that found with a random distribution of objects. That is, we would observe a `flatter' distribution in Fig. \ref{fig:triggering} compared to the randomizations. Clearly, Fig. \ref{fig:triggering} shows that massive star formation is significantly boosted in clouds found within 10 pc of a SC. The same result is also apparent by comparing the histograms of w/ SC and w/o SC samples. At $R_\m{s}$ = 10, we find a clear dichotomy between both samples, where the w/ SC sample are shown to contain many more sources than the w/o SC sample. This dichotomy is most pronounced for the MYSOs and VMYSOs. More specifically, we find that with $R_\m{s}$ = 10, $\sim$\,65\%/90\% of the clouds in the w/o SC sample ($d$\,$\textgreater$\,10 pc) are devoid of any MYSO/VMYSOs, whereas this fraction is only $\sim$\,15\%/45\% for the clouds in the w/ SC sample ($d$\,$\textless$\,10 pc). We conclude that clouds within 10 pc of a SC have much higher MYSO/VMYSO (and DC) number counts, implying a correlation between the presence of a SC and an increased rate of massive star formation over the past $\sim$\,10$^5$ yr.

By increasing our search radius to $R_\m{s}$ = 30 pc and $R_\m{s}$ = 100 pc, the number of clouds in the w/ SC sample eventually exceeds that of the w/o SC sample. We find that by increasing $R_\m{s}$, the dichotomy in number counts between the w/ SC and w/o SC sample disappears, causing the histograms in Fig. \ref{fig:triggering} to converge. This implies that the correlation between the presence of SCs and the rate of massive star formation becomes less pronounced at larger distances.

Table \ref{tab:yield} shows that VMYSOs are almost exclusively found within 30 pc of a SC. A similar trend is seen for DCs, which again may imply that DCs represent a part of the VMYSO population (Sec. \ref{sec:distribution}). Thus, the connection between different generations of massive stars may be stronger for O-type progenitor stars (VMYSOs; Sec. \ref{sec:distribution}) than B-type progenitor stars (MYSOs). Finally, note that 35\% of $N_\m{max}$ are not found within $R_\m{s}$ = 100 pc, even though we established that the median radius of islands and clouds are 28.4 pc and 17.0 pc, respectively. This results from the fact that many GMCs are far from spherical (Fig. \ref{fig:postage}) and may better be represented by a filamentary-like morphology.

Figure \ref{fig:triggering} shows that the amount of MYSOs/VMYSOs may vary greatly between the w/SCs and w/o SCs cloud samples. However, the associated mass ($M_\m{cloud}$) and average surface density ($\mean{\Sigma_\m{cloud}}$) distributions of both samples (Fig. \ref{fig:triggering}) are remarkably similar. Thus, the rate of massive star formation in clouds near SCs does not appear to correlate strongly with these specific cloud properties.

\section{Discussion}\label{sec:discussion}

In this work we have presented the study of a unique dataset that offers a galaxy-wide view of molecular clouds ($M$\,$\geq$\,3\,$\times$\,10$^4$ M$_\m{\odot}$), young ($\sim$\,10$^5$ yr) sources on their way of becoming massive stars ($M$\,$\textgreater$\,8 M$_\m{\odot}$), and young ($\textless$\,10 Myr) optical stellar clusters. The sheer size of the data set allowed us to identify the location, clustering, and follow the propagation of massive star formation in giant molecular clouds.

In the LMC, massive stars do not typically form at the highest column densities nor centre of their parent GMCs at the $\sim$\,6 pc resolution of our observations (Figs. \ref{fig:postage} and \ref{fig:histogram}). Half of our sample of MYSOs, VMYSOs, and DCs are formed $\sim$\,10 pc away from local column density peaks (Fig. \ref{fig:cumulative}). Massive star formation clusters over different generations and on scales much smaller than the parent molecular cloud (Fig. \ref{fig:clustering}), regardless if we include the diffuse parts of the GMCs (`islands') or focus on the highest column density structures alone (`clouds'). While the rate of massive star formation is significantly boosted in clouds near SCs (Fig. \ref{fig:triggering}), comparison of molecular clouds associated with SCs with those that are not reveals no significant difference in total mass and average surface density.

\subsection{The location of massive star formation in GMCs}\label{sec:loc}

The dearth of MYSOs at high column densities in GMCs (Sec. \ref{sec:distribution}) merits further discussion. We have ruled out if completeness systematically affects our analysis (Sec. \ref{sec:completeness}). Alternatively, feedback from massive stars can dynamically alter the cloud material, which may lead to an apparent offset between young massive stars and high-column density material. However, given the estimated age of our sample of YSOs ($\sim$\,10$^5$ yr; Sec. \ref{sec:catalogue}), it is unlikely that we are tracing feedback processes on the physical scales we probe ($\gtrsim$\,6 pc; Sec. \ref{sec:column}). In fact, it is unclear when MYSOs start ionizing their surroundings \citep{churchwell_2002,hoare_2007}: current galaxy-wide LMC radio maps of free-free emission \citep{dickel_2005,hughes_2007} do not have the angular resolution to asses if some of our MYSO/VMYSO/DC sources have reached the ultra-compact \HII\ region phase. Even if we assume that our sources have started ionizing their surroundings, analytical solutions \citep{spitzer_1978, dyson_1980}, 1D simulations \citep{raga_2012}, and turbulent 3D simulations \citep{tremblin_2014} reveal that \HII\ regions in typical molecular cloud conditions only reach sizes of $\lesssim$\,0.5 pc within $10^5$ yr, which is small compared to the resolution of our column density maps ($\sim$\,6 pc). Thus, the timescales involved are incompatible with our young massive stars having created several parsec-sized cavities within their natal clouds. Indeed, high-resolution H$\alpha$ imaging (Fig. \ref{fig:postage}) do not show the indications of large-scale feedback processes, confirming the embedded nature of our MYSO sample.

As a caveat, we note that the presence of internal heating sources can elevate the local dust temperature, overestimating the mass-averaged temperature along the line-of-sight, and thereby underestimating the total observed column density. Conversely, in the absence of an internal heating source, the observed SED will be biased towards the irradiated outskirts of  clouds or cores as opposed to the dark, cold cores containing the bulk of the mass. These effects are inherent to FIR SED fitting, and have been addressed in many studies \citep[e.g.,][]{malinen_2011,ysard_2012}. The underestimation of mass appears to be larger for starless than protostellar cores \citep{malinen_2011}, because internal heating renders dust more easily visible and estimations of cloud masses become more reliable. As noted by \citet{juvela_2013}, quantifying the extent to which our column densities are affected by the absence/presence of, e.g., YSOs could only be alleviated by knowledge of the temperature structure of the source and the detailed structure of the molecular cloud (i.e., density and line-of-sight depth). Alternatively, we can resort to gas-based column density tracers: the MAGMA CO data show very similar distributions (Fig. \ref{fig:postage}), however we have already pointed out that $^{12}$CO\,(1-0) has observational limitations on its own (Sec. \ref{sec:mapandcloud}). Future observations of GMCs in various tracers of different critical densities will allow to map the internal structure of GMCs at a high dynamical range. In this way, we will be able to quantify whether our FIR-derived column density maps are significantly affected by line-of-sight temperature gradients, and how this impacts our results on the location of MYSOs in GMCs (Sec. \ref{sec:distribution}).

\subsubsection{Comparison with Galactic studies}\label{sec:galactic}

Galactic studies of (massive) star formation are traditionally complicated by confusion, large angular scales, and distance ambiguity. Our study targeting massive star formation in the LMC circumvents these limitations and, as a result, there are (as of yet) no Milky Way studies of similar size, combining observations of hundreds of GMCs, associated stellar clusters, and a complete census of embedded massive star formation over the past $\sim$\,10$^5$ yr, which allows for a statistical study of massive star formation and its dependence on environment and evolutionary state. 

In the Galaxy, massive stars may form within IRDCs \citep[e.g.,][]{beuther_2005, rathborne_2006}. One might question if the HERITAGE maps offer sufficient spatial resolution to be sensitive to typical IRDCs and star forming clumps such as those seen in (nearby) Galactic clouds. Surely, on small scales the column densities at the position of the MYSOs may be very high, and beam dilution may render these sites undetectable to our observations. IRDCs have typical sizes of $\sim$\,5 pc \citep{simon_2006}, while massive star forming clumps have sizes of order $\sim$\,1 pc \citep{tan_2014}. Both of these structures would be unresolved at the resolution of our $N$\,(H$_2$) map ($\sim$\,6 pc; Sec. \ref{sec:mapandcloud}). In this respect, a proper exercise is to consider if we would be able to detect the nearest example of massive star formation, located within the Orion A molecular cloud. Orion A has a surface area of $\sim$\,2200 pc$^2$ and contains $\sim$\,10$^5$ M$_\odot$ of molecular mass \citep{wilson_2005}, large and massive enough to be resolved and detected in the HERITAGE maps. On the northern tip of the cloud lies the `integral-shaped filament' (ISF; \citealt{bally_1987}), a dense ridge (9\,$\times$\,0.5 pc) containing $\sim$\,5\,$\times$10$^3$ M$_\m{\odot}$ of molecular gas \citep{bally_1987,berne_2014}. While the ISF only comprises 1/500 of the surface area of Orion A \citep{wilson_2005}, it contains 1/20 of its mass, thereby locally increasing the column density by a factor of $\sim$\,25. If we assume that Orion A (cloud) and the ISF (filament) are hierarchically perched on top of one another \citep{wilson_2005}, we can estimate the observed column density contrast of the Orion ISF region with respect to the entire Orion A cloud, taking into account the beam-filling factor $f$ of the ISF in our 6 pc resolution maps ($f_\m{ISF}$\,$\sim$\,0.1). We write the column density contrast as $\eta$ = ($N_\m{H_2}\m{[Ori\,A]}$ + $f_\m{ISF}N_\m{H_2}\m{[ISF]}$)/($N_\m{H_2}\m{[Ori\,A]}$) $\sim$ 3.5. We conclude that the ISF, and thereby a site of massive star formation like Orion, should be detectable in our column density maps of the LMC. 

The above derivation illustrates a key point: massive stars in the LMC do not appear to form in environments with masses similar to that of Galactic IRDCs ($\sim$\,5\,$\times$\,10$^3$ $M_\odot$; \citealt{simon_2006}). Still, we would expect massive stars to form in density enhancements unresolved at our $\sim$\,6 pc resolution, consistent with gas clumps and massive cores observed in the Galaxy ($\lesssim$\,1 pc; \citealt{tan_2014}). Our results illustrate that the clumps and cores forming massive stars are created outside of the densest, most opaque regions of GMCs (Figs. \ref{fig:postage}, \ref{fig:histogram}, and \ref{fig:cumulative}). To estimate an upper limit for the mass of these systems, we use the median surface density of all GMCs in the LMC, equalling 23 M$_{\odot}$ pc$^{-2}$ and 37 M$_{\odot}$ pc$^{-2}$ for islands and clouds, respectively. At these surface densities and assuming a Gaussian beam with FWHM = 6 pc, a compact (unresolved) gas clump of mass $M_\m{cl}$ = 500 M$_\m{\odot}$, massive enough to form a cluster of mass $M_\star$ containing a maximum stellar mass of $\sim$\,25 $M_\m{\odot}$, would lead to a column density contrast $\eta$\,$\sim$\,1.5 (assuming an efficiency of $M_\m{cl}$/$M_\star$$\sim$\,0.5 and a Kroupa initial mass function; \citealt{tan_2014,kroupa_2001}). Such a column density contrast may be confused with the background GMC. Alternatively, on the scales of individual gas clumps and massive cores ($\lesssim$\,1 pc), the MYSOs/VMYSOs/DCs may simply have destroyed their natal star forming clump, given that the current estimated destruction timescale ($\sim$\,3\,$\times$\,10$^5$ yr; \citealt{seale_2012}) is of order of the estimated age of our MYSO sample ($\sim$\,10$^5$ yr; Sec. \ref{sec:catalogue}).

We conclude that massive stars in the LMC appear to form in clumps with masses up to $\lesssim$\,500 M$_\m{\odot}$, which is consistent with current theories and observations of massive star formation (see \citealt{tan_2014} for a recent comprehensive review). However, the observation that massive stars (and, presumably, their natal clumps) form outside of the main body of molecular gas in GMCs is puzzling, and may provide important clues to the collapse of molecular clouds and the initial conditions that may lead to the formation of massive stars (see Sec. \ref{sec:modes}). We note that these results may also apply to massive star formation in the Galaxy, given that most Galactic IRDCs show no sign of active star formation \citep{chambers_2009}, while the recently discovered `giant molecular filaments' in the Galaxy \citep{jackson_2010, ragan_2014} reveal many massive cores and ultra-compact \HII\ regions around the edges of the giant filaments \citep{abreu-vicente_2016}. 

We argue that in order to advance our understanding on the location and formation of massive stars within GMCs, it is essential to consider entire GMC complexes (including the potential influence of external factors; Sec. \ref{sec:modes}), instead of merely focusing on `hot spots' that appear to be prime candidates for the formation of massive stars (i.e., IRDCs). In this regard, high-resolution follow-up observations of GMCs complexes with the Atacama Large Millimeter Array \citep[e.g.,][]{indebetouw_2013, fukui_2015, nayak_2016} together with sensitive observations of MYSOs with the James Webb Space Telescope will provide suitable tools to advance our understanding on the location, clustering, and propagation of massive star formation and its relation to the large-scale structure of GMCs and the ISM of galaxies.

\subsection{The clustering and propagation of massive star formation in GMCs}\label{sec:clustering}
 
Figure \ref{fig:triggering} revealed that there is a strong dichotomy in massive star formation rate between Type 3 clouds, depending on their location from SCs. Specifically, massive star formation is significantly boosted in clouds near SCs, with the effect becoming less pronounced at larger distances from SCs (Fig. \ref{fig:triggering}). The results suggest a connection between different generations of massive stars on timescales up to 10 Myr. It is tempting to take this result as evidence for triggered star formation where, once star formation is initiated, the interaction of the newly formed massive stars with their environment drives the formation of the next generation (see Sec. \ref{sec:modes}). 

One may argue that massive star formation can exclusively be found in regions with certain physical conditions (e.g., above a mass or density threshold) and that it is only natural to find massive stars clustered in these particular regions. After all, it is very well known that massive stars form almost exclusively in clustered environments \citep{ladalada_2003}. However, the rate of massive star formation in clouds near SCs does not appear to correlate with the {\em global} properties $M_\m{cloud}$ and $\mean{\Sigma_\m{cloud}}$ (Fig. \ref{fig:triggering}). This may be related to the results from Sec. \ref{sec:results}: massive star formation takes place on scales much smaller than islands (Tab. \ref{tab:mysos}) as well as clouds (Fig. \ref{fig:clustering}). These results indicate that massive star formation is a {\em local} process within GMCs. Massive star formation as a local process would disconnect the rate of massive star formation from the global cloud properties $M_\m{cloud}$ and $\mean{\Sigma_\m{cloud}}$. We note that similar results were obtained for nearby molecular clouds \citep{lada_2010,heiderman_2010}, where star formation appears poorly correlated with total molecular cloud mass, but is instead closely related to the dense gas fraction within molecular clouds. However, it is important to note that these results were based on low-mass star formation: it is unclear if the same laws apply to high-mass star formation (Sec. \ref{sec:modes}). Unfortunately, our observations do not have the resolution to discern between dense and diffuse gas within the GMCs: higher-resolution data resolving the intrinsic GMC structure is needed to discern if low-mass and high-mass stars form alike, or if they form through different pathways.

\subsection{The modes of massive star formation \& comparison with earlier works}\label{sec:modes}

At this point, we reiterate the two main results presented in this work. First, MYSOs do not form at column density peaks of GMCs. Second, massive star formation is more active in clouds close to young SCs. As noted in Sec. \ref{sec:galactic}, these results may provide important clues to the collapse of molecular clouds and the initial conditions that lead to the formation of massive stars. Below, we explore routes to the formation of massive stars identified in the literature that may be consistent with our observations.

Numerical and analytical studies have identified mechanisms that can lead to the formation of massive cores {\em on the edges} of molecular clouds \citep{burkert_2004,heitsch_2008,pon_2011,li_2016}. For example, `edge effects' arise in collapsing finite cloud sheets, where material accumulates and fragments at the outer boundaries of the cloud where the gravitational acceleration is greatest \citep{burkert_2004}. In addition, \citet{heitsch_2008} studied the formation of molecular clouds in large-scale colliding flows, and showed that while {\em global} collapse of a molecular cloud creates centrally located large-scale filaments, {\em local} gravitational collapse can lead to high-mass cores far away from the centers of molecular clouds on timescales much shorter than the global dynamical collapse time. 

The above described mechanisms can lead to the `spontaneous' formation of massive stars at the outskirts of molecular clouds. A different train of thought involves the notion that high-mass star formation has to be induced or `triggered' as opposed to their lower-mass counterparts \citep{shu_1987}. The actual driving agents of triggering may vary and act on a wide range of different scales, from galaxy mergers \citep{woods_2006}, galaxy-scale turbulence \citep{mac-low_2004}, spiral arm passages \citep{roberts_1969}, supershells \citep{tenorio-tagle_1988}, cloud-cloud collisions \citep{scoville_1986,fukui_2015}, to that of single stars or clusters through `cloud-crushing' \citep{bertoldi_1989} or the `collect-and-collapse' process \citep{elmegreen_1977,zavagno_2007}. While studies of individual, isolated regions such as RCW 120 \citep[e.g.,][]{zavagno_2010} have unambiguously demonstrated the importance of triggered star formation, its relevance on a larger scale has remained controversial \citep{dale_2015}. The controversy arises largely because most regions of massive star formation show lots of star formation related activity in different stages of evolution, and that makes pinpointing the effects of triggered star formation difficult.  

Local gravitational collapse of GMCs as a mode of massive star formation may explain the distribution of young massive stars in the LMC (Fig. \ref{fig:histogram}), as this would not {\em a priori} favor the central (i.e., highest column density) regions within GMCs as the principle formation site of massive stars. Alternatively, feedback from massive stars may trigger the formation of a next generation in regions of a GMC that may not necessarily correlate with total column density. Both of these scenarios can occur within localized (small) regions of GMCs, consistent with the scale size of clustering versus that of the size of GMCs (Fig. \ref{fig:clustering}), and the rate of massive star formation disconnected to the global properties of GMCs (Fig. \ref{fig:triggering}). Moreover, triggering as a key mode for massive star formation links different generations of massive stars and would explain the strong correlation between the presence of SCs and the rate of massive star formation (Fig. \ref{fig:triggering}). While the importance of induced star formation has been a subject of debate for decades \citep{shu_1987,elmegreen_1998,dale_2015}, it appears consistent with results on individual (nearby) massive star forming regions in the Galaxy \citep{blaauw_1964, elmegreen_1977, povich_2009, zavagno_2010}, and we argue that the close association of exposed clusters with nearby embedded massive stars provides further support for the importance of triggered star formation on a galaxy-wide scale.

\section{Conclusions}\label{sec:conclusions}

We have studied massive star formation in GMCs of the LMC using an unbiased sample of $\sim$\,700 MYSOs, $\sim$\,200 GMCs, and $\sim$\,100 SCs. Unhindered by confusion or luminosity uncertainties that typically hamper Galactic studies, we were able to study the location, clustering, and propagation of massive star formation within GMCs. Our main results are as follows:

\begin{enumerate}

\item[-] Our MYSO catalogue is complete for Stage 1 MYSOs of mass $M$\,$\textgreater$\,8 $M_\m{\odot}$, provided that they have mid-IR counterparts (Sec. \ref{sec:completeness}).

\item[-] We find ongoing massive star formation (i.e., over the past $\sim$\,10$^5$ yr) in 33\% or 48\% of the LMC GMCs, depending if we consider `clouds' or `islands' (Sec. \ref{sec:dendrogram}). We substantiate the classification scheme from \citet{kawamura_2009} by revealing that Type 1 GMCs are (mostly) devoid of massive star formation (Tab. \ref{tab:mysos}). 

\item[-] We find that massive stars do not form at the peak column densities within GMCs at the $\sim$\,6 pc resolution of our observations (Figs. \ref{fig:postage} and \ref{fig:histogram}). Specifically, half of our sample of MYSOs/VMYSOs/DCs are located $\textgreater$\,10 pc from $N_\m{max}$ (Fig. \ref{fig:cumulative}). We have excluded completeness or feedback as a cause for this result (Sec. \ref{sec:results}). 

\item[-] By means of angular correlation functions (Eqs. \ref{eq:auto} \& \ref{eq:cross}; Fig. \ref{fig:clustering}), we have demonstrated that MYSOs/VMYSOs/DCs are strongly clustered on scales much smaller than the size of CO islands and clouds. The auto-correlations show very similar results compared to their respective cross-correlations with SCs, indicating that massive star formation is clustered over different generations on timescales up to 10 Myr.

\item[-] We find that the rate of massive star formation is significantly elevated in clouds near SCs (Fig. \ref{fig:triggering}). At the same time, the rate of massive star formation in these clouds appears unrelated to the global cloud properties $M_\m{cloud}$ and $\mean{\Sigma_\m{cloud}}$. The relative increase in massive star formation becomes less pronounced at larger distances from the SCs.

\end{enumerate}

We argue that massive star formation is a local process within GMCs. It appears that the initial conditions leading to massive star formation do not necessarily occur in the densest, most opaque regions of GMCs. Our results reveal a close connection between different generations of massive stars on timescales up to 10 Myr, which may provide further support for triggering as a key mode for massive star formation, which in its turn could proceed very differently compared to their lower mass counterparts.
 
\acknowledgments The authors wish to thank Kirill Tchernyshyov and Karl Gordon for many useful discussions, and Henrik Beuther and Sarah Kendrew for comments on an (early) draft of this paper. We thank the MAGMA team for permission to use the DR3 products in advance of publication. BO is supported through NASA ADAP grant NNX15AF17G. 

\bibliographystyle{aa} 
\bibliography{1_msf} 

\begin{thebibliography}{96}
\expandafter\ifx\csname natexlab\endcsname\relax\def\natexlab#1{#1}\fi

\bibitem[{{Abreu-Vicente} {et~al.}(2016){Abreu-Vicente}, {Ragan},
  {Kainulainen}, {Henning}, {Beuther}, \& {Johnston}}]{abreu-vicente_2016}
{Abreu-Vicente}, J., {Ragan}, S., {Kainulainen}, J., {et~al.} 2016, \aap, 590,
  A131

\bibitem[{{Andr{\'e}} {et~al.}(2014){Andr{\'e}}, {Di Francesco},
  {Ward-Thompson}, {Inutsuka}, {Pudritz}, \& {Pineda}}]{andre_2014}
{Andr{\'e}}, P., {Di Francesco}, J., {Ward-Thompson}, D., {et~al.} 2014,
  Protostars and Planets VI, 27

\bibitem[{{Andr{\'e}} {et~al.}(2010){Andr{\'e}}, {Men'shchikov}, {Bontemps},
  {K{\"o}nyves}, {Motte}, {Schneider}, {Didelon}, {Minier}, {Saraceno},
  {Ward-Thompson}, {di Francesco}, {White}, {Molinari}, {Testi}, {Abergel},
  {Griffin}, {Henning}, {Royer}, {Mer{\'{\i}}n}, {Vavrek}, {Attard},
  {Arzoumanian}, {Wilson}, {Ade}, {Aussel}, {Baluteau}, {Benedettini},
  {Bernard}, {Blommaert}, {Cambr{\'e}sy}, {Cox}, {di Giorgio}, {Hargrave},
  {Hennemann}, {Huang}, {Kirk}, {Krause}, {Launhardt}, {Leeks}, {Le Pennec},
  {Li}, {Martin}, {Maury}, {Olofsson}, {Omont}, {Peretto}, {Pezzuto}, {Prusti},
  {Roussel}, {Russeil}, {Sauvage}, {Sibthorpe}, {Sicilia-Aguilar}, {Spinoglio},
  {Waelkens}, {Woodcraft}, \& {Zavagno}}]{andre_2010}
{Andr{\'e}}, P., {Men'shchikov}, A., {Bontemps}, S., {et~al.} 2010, \aap, 518,
  L102

\bibitem[{{Astropy Collaboration} {et~al.}(2013){Astropy Collaboration},
  {Robitaille}, {Tollerud}, {Greenfield}, {Droettboom}, {Bray}, {Aldcroft},
  {Davis}, {Ginsburg}, {Price-Whelan}, {Kerzendorf}, {Conley}, {Crighton},
  {Barbary}, {Muna}, {Ferguson}, {Grollier}, {Parikh}, {Nair}, {Unther},
  {Deil}, {Woillez}, {Conseil}, {Kramer}, {Turner}, {Singer}, {Fox}, {Weaver},
  {Zabalza}, {Edwards}, {Azalee Bostroem}, {Burke}, {Casey}, {Crawford},
  {Dencheva}, {Ely}, {Jenness}, {Labrie}, {Lim}, {Pierfederici}, {Pontzen},
  {Ptak}, {Refsdal}, {Servillat}, \& {Streicher}}]{robitaille_2013}
{Astropy Collaboration}, {Robitaille}, T.~P., {Tollerud}, E.~J., {et~al.} 2013,
  \aap, 558, A33

\bibitem[{{Bally} {et~al.}(1987){Bally}, {Langer}, {Stark}, \&
  {Wilson}}]{bally_1987}
{Bally}, J., {Langer}, W.~D., {Stark}, A.~A., \& {Wilson}, R.~W. 1987, \apjl,
  312, L45

\bibitem[{{Bernard} {et~al.}(2008){Bernard}, {Reach}, {Paradis}, {Meixner},
  {Paladini}, {Kawamura}, {Onishi}, {Vijh}, {Gordon}, {Indebetouw}, {Hora},
  {Whitney}, {Blum}, {Meade}, {Babler}, {Churchwell}, {Engelbracht}, {For},
  {Misselt}, {Leitherer}, {Cohen}, {Boulanger}, {Frogel}, {Fukui}, {Gallagher},
  {Gorjian}, {Harris}, {Kelly}, {Latter}, {Madden}, {Markwick-Kemper},
  {Mizuno}, {Mizuno}, {Mould}, {Nota}, {Oey}, {Olsen}, {Panagia},
  {Perez-Gonzalez}, {Shibai}, {Sato}, {Smith}, {Staveley-Smith}, {Tielens},
  {Ueta}, {Van Dyk}, {Volk}, {Werner}, \& {Zaritsky}}]{bernard_2008}
{Bernard}, J.-P., {Reach}, W.~T., {Paradis}, D., {et~al.} 2008, \aj, 136, 919

\bibitem[{{Bern{\'e}} {et~al.}(2014){Bern{\'e}}, {Marcelino}, \&
  {Cernicharo}}]{berne_2014}
{Bern{\'e}}, O., {Marcelino}, N., \& {Cernicharo}, J. 2014, \apj, 795, 13

\bibitem[{{Bertoldi}(1989)}]{bertoldi_1989}
{Bertoldi}, F. 1989, \apj, 346, 735

\bibitem[{{Beuther} {et~al.}(2005){Beuther}, {Sridharan}, \&
  {Saito}}]{beuther_2005}
{Beuther}, H., {Sridharan}, T.~K., \& {Saito}, M. 2005, \apjl, 634, L185

\bibitem[{{Bica} {et~al.}(1996){Bica}, {Claria}, {Dottori}, {Santos}, \&
  {Piatti}}]{bica_1996}
{Bica}, E., {Claria}, J.~J., {Dottori}, H., {Santos}, Jr., J.~F.~C., \&
  {Piatti}, A.~E. 1996, \apjs, 102, 57

\bibitem[{{Blaauw}(1964)}]{blaauw_1964}
{Blaauw}, A. 1964, \araa, 2, 213

\bibitem[{{Bolatto} {et~al.}(2013){Bolatto}, {Wolfire}, \&
  {Leroy}}]{bolatto_2013}
{Bolatto}, A.~D., {Wolfire}, M., \& {Leroy}, A.~K. 2013, \araa, 51, 207

\bibitem[{{Bradshaw} {et~al.}(2011){Bradshaw}, {Almaini}, {Hartley}, {Chuter},
  {Simpson}, {Conselice}, {Dunlop}, {McLure}, \& {Cirasuolo}}]{bradshaw_2011}
{Bradshaw}, E.~J., {Almaini}, O., {Hartley}, W.~G., {et~al.} 2011, \mnras, 415,
  2626

\bibitem[{{Burkert} \& {Hartmann}(2004)}]{burkert_2004}
{Burkert}, A. \& {Hartmann}, L. 2004, \apj, 616, 288

\bibitem[{{Chambers} {et~al.}(2009){Chambers}, {Jackson}, {Rathborne}, \&
  {Simon}}]{chambers_2009}
{Chambers}, E.~T., {Jackson}, J.~M., {Rathborne}, J.~M., \& {Simon}, R. 2009,
  \apjs, 181, 360

\bibitem[{{Churchwell}(2002)}]{churchwell_2002}
{Churchwell}, E. 2002, \araa, 40, 27

\bibitem[{{Dale} {et~al.}(2015){Dale}, {Haworth}, \& {Bressert}}]{dale_2015}
{Dale}, J.~E., {Haworth}, T.~J., \& {Bressert}, E. 2015, \mnras, 450, 1199

\bibitem[{{Dickel} {et~al.}(2005){Dickel}, {McIntyre}, {Gruendl}, \&
  {Milne}}]{dickel_2005}
{Dickel}, J.~R., {McIntyre}, V.~J., {Gruendl}, R.~A., \& {Milne}, D.~K. 2005,
  \aj, 129, 790

\bibitem[{{Dunham} {et~al.}(2014){Dunham}, {Stutz}, {Allen}, {Evans},
  {Fischer}, {Megeath}, {Myers}, {Offner}, {Poteet}, {Tobin}, \&
  {Vorobyov}}]{dunham_2014}
{Dunham}, M.~M., {Stutz}, A.~M., {Allen}, L.~E., {et~al.} 2014, Protostars and
  Planets VI, 195

\bibitem[{{Dupac} {et~al.}(2003){Dupac}, {Bernard}, {Boudet}, {Giard},
  {Lamarre}, {M{\'e}ny}, {Pajot}, {Ristorcelli}, {Serra}, {Stepnik}, \&
  {Torre}}]{dupac_2003}
{Dupac}, X., {Bernard}, J.-P., {Boudet}, N., {et~al.} 2003, \aap, 404, L11

\bibitem[{{Dyson} \& {Williams}(1980)}]{dyson_1980}
{Dyson}, J.~E. \& {Williams}, D.~A. 1980, {Physics of the interstellar medium}

\bibitem[{{Elmegreen}(1998)}]{elmegreen_1998}
{Elmegreen}, B.~G. 1998, in Astronomical Society of the Pacific Conference
  Series, Vol. 148, Origins, ed. C.~E. {Woodward}, J.~M. {Shull}, \& H.~A.
  {Thronson}, Jr., 150

\bibitem[{{Elmegreen} \& {Lada}(1977)}]{elmegreen_1977}
{Elmegreen}, B.~G. \& {Lada}, C.~J. 1977, \apj, 214, 725

\bibitem[{{Evans} {et~al.}(2009){Evans}, {Dunham}, {J{\o}rgensen}, {Enoch},
  {Mer{\'{\i}}n}, {van Dishoeck}, {Alcal{\'a}}, {Myers}, {Stapelfeldt},
  {Huard}, {Allen}, {Harvey}, {van Kempen}, {Blake}, {Koerner}, {Mundy},
  {Padgett}, \& {Sargent}}]{evans_2009}
{Evans}, II, N.~J., {Dunham}, M.~M., {J{\o}rgensen}, J.~K., {et~al.} 2009,
  \apjs, 181, 321

\bibitem[{{Fukui} {et~al.}(2015){Fukui}, {Harada}, {Tokuda}, {Morioka},
  {Onishi}, {Torii}, {Ohama}, {Hattori}, {Nayak}, {Meixner}, {Sewi{\l}o},
  {Indebetouw}, {Kawamura}, {Saigo}, {Yamamoto}, {Tachihara}, {Minamidani},
  {Inoue}, {Madden}, {Galametz}, {Lebouteiller}, {Mizuno}, \&
  {Chen}}]{fukui_2015}
{Fukui}, Y., {Harada}, R., {Tokuda}, K., {et~al.} 2015, \apjl, 807, L4

\bibitem[{{Fukui} {et~al.}(2008){Fukui}, {Kawamura}, {Minamidani}, {Mizuno},
  {Kanai}, {Mizuno}, {Onishi}, {Yonekura}, {Mizuno}, {Ogawa}, \&
  {Rubio}}]{fukui_2008}
{Fukui}, Y., {Kawamura}, A., {Minamidani}, T., {et~al.} 2008, \apjs, 178, 56

\bibitem[{{Goodman} {et~al.}(2009){Goodman}, {Pineda}, \&
  {Schnee}}]{goodman_2009}
{Goodman}, A.~A., {Pineda}, J.~E., \& {Schnee}, S.~L. 2009, \apj, 692, 91

\bibitem[{{Gordon} {et~al.}(2014){Gordon}, {Roman-Duval}, {Bot}, {Meixner},
  {Babler}, {Bernard}, {Bolatto}, {Boyer}, {Clayton}, {Engelbracht}, {Fukui},
  {Galametz}, {Galliano}, {Hony}, {Hughes}, {Indebetouw}, {Israel}, {Jameson},
  {Kawamura}, {Lebouteiller}, {Li}, {Madden}, {Matsuura}, {Misselt}, {Montiel},
  {Okumura}, {Onishi}, {Panuzzo}, {Paradis}, {Rubio}, {Sandstrom}, {Sauvage},
  {Seale}, {Sewi{\l}o}, {Tchernyshyov}, \& {Skibba}}]{gordon_2014}
{Gordon}, K.~D., {Roman-Duval}, J., {Bot}, C., {et~al.} 2014, \apj, 797, 85

\bibitem[{{Gruendl} \& {Chu}(2009)}]{gruendl_2009}
{Gruendl}, R.~A. \& {Chu}, Y.-H. 2009, \apjs, 184, 172

\bibitem[{{Heiderman} {et~al.}(2010){Heiderman}, {Evans}, {Allen}, {Huard}, \&
  {Heyer}}]{heiderman_2010}
{Heiderman}, A., {Evans}, II, N.~J., {Allen}, L.~E., {Huard}, T., \& {Heyer},
  M. 2010, \apj, 723, 1019

\bibitem[{{Heitsch} {et~al.}(2008){Heitsch}, {Hartmann}, {Slyz}, {Devriendt},
  \& {Burkert}}]{heitsch_2008}
{Heitsch}, F., {Hartmann}, L.~W., {Slyz}, A.~D., {Devriendt}, J.~E.~G., \&
  {Burkert}, A. 2008, \apj, 674, 316

\bibitem[{{Hoare} \& {Franco}(2007)}]{hoare_2007}
{Hoare}, M.~G. \& {Franco}, J. 2007, Astrophysics and Space Science
  Proceedings, 1, 61

\bibitem[{{Hughes} {et~al.}(2013){Hughes}, {Meidt}, {Colombo}, {Schinnerer},
  {Pety}, {Leroy}, {Dobbs}, {Garc{\'{\i}}a-Burillo}, {Thompson}, {Dumas},
  {Schuster}, \& {Kramer}}]{hughes_2013}
{Hughes}, A., {Meidt}, S.~E., {Colombo}, D., {et~al.} 2013, \apj, 779, 46

\bibitem[{{Hughes} {et~al.}(2007){Hughes}, {Staveley-Smith}, {Kim}, {Wolleben},
  \& {Filipovi{\'c}}}]{hughes_2007}
{Hughes}, A., {Staveley-Smith}, L., {Kim}, S., {Wolleben}, M., \&
  {Filipovi{\'c}}, M. 2007, \mnras, 382, 543

\bibitem[{{Indebetouw} {et~al.}(2013){Indebetouw}, {Brogan}, {Chen}, {Leroy},
  {Johnson}, {Muller}, {Madden}, {Cormier}, {Galliano}, {Hughes}, {Hunter},
  {Kawamura}, {Kepley}, {Lebouteiller}, {Meixner}, {Oliveira}, {Onishi}, \&
  {Vasyunina}}]{indebetouw_2013}
{Indebetouw}, R., {Brogan}, C., {Chen}, C.-H.~R., {et~al.} 2013, \apj, 774, 73

\bibitem[{{Jackson} {et~al.}(2010){Jackson}, {Finn}, {Chambers}, {Rathborne},
  \& {Simon}}]{jackson_2010}
{Jackson}, J.~M., {Finn}, S.~C., {Chambers}, E.~T., {Rathborne}, J.~M., \&
  {Simon}, R. 2010, \apjl, 719, L185

\bibitem[{{Jameson} {et~al.}(2016){Jameson}, {Bolatto}, {Leroy}, {Meixner},
  {Roman-Duval}, {Gordon}, {Hughes}, {Israel}, {Rubio}, {Indebetouw}, {Madden},
  {Bot}, {Hony}, {Cormier}, {Pellegrini}, {Galametz}, \&
  {Sonneborn}}]{jameson_2015}
{Jameson}, K.~E., {Bolatto}, A.~D., {Leroy}, A.~K., {et~al.} 2016, \apj, 825,
  12

\bibitem[{{Juvela} {et~al.}(2013){Juvela}, {Malinen}, \&
  {Lunttila}}]{juvela_2013}
{Juvela}, M., {Malinen}, J., \& {Lunttila}, T. 2013, \aap, 553, A113

\bibitem[{{Kawamura} {et~al.}(2009){Kawamura}, {Mizuno}, {Minamidani},
  {Filipovi{\'c}}, {Staveley-Smith}, {Kim}, {Mizuno}, {Onishi}, {Mizuno}, \&
  {Fukui}}]{kawamura_2009}
{Kawamura}, A., {Mizuno}, Y., {Minamidani}, T., {et~al.} 2009, \apjs, 184, 1

\bibitem[{{Kendrew}(2015)}]{kendrew_code_2015}
{Kendrew}, S. 2015, {milkywayproject\_triggering: Correlation functions for two
  catalog datasets}, Astrophysics Source Code Library

\bibitem[{{Kendrew} {et~al.}(2016){Kendrew}, {Beuther}, {Simpson}, {Csengeri},
  {Wienen}, {Lintott}, {Povich}, {Beaumont}, \& {Schuller}}]{kendrew_2016}
{Kendrew}, S., {Beuther}, H., {Simpson}, R., {et~al.} 2016, \apj, 825, 142

\bibitem[{{Kendrew} {et~al.}(2012){Kendrew}, {Simpson}, {Bressert}, {Povich},
  {Sherman}, {Lintott}, {Robitaille}, {Schawinski}, \&
  {Wolf-Chase}}]{kendrew_2012}
{Kendrew}, S., {Simpson}, R., {Bressert}, E., {et~al.} 2012, \apj, 755, 71

\bibitem[{{Kenyon} {et~al.}(1990){Kenyon}, {Hartmann}, {Strom}, \&
  {Strom}}]{kenyon_1990}
{Kenyon}, S.~J., {Hartmann}, L.~W., {Strom}, K.~M., \& {Strom}, S.~E. 1990,
  \aj, 99, 869

\bibitem[{{K{\"o}nyves} {et~al.}(2010){K{\"o}nyves}, {Andr{\'e}},
  {Men'shchikov}, {Schneider}, {Arzoumanian}, {Bontemps}, {Attard}, {Motte},
  {Didelon}, {Maury}, {Abergel}, {Ali}, {Baluteau}, {Bernard}, {Cambr{\'e}sy},
  {Cox}, {di Francesco}, {di Giorgio}, {Griffin}, {Hargrave}, {Huang}, {Kirk},
  {Li}, {Martin}, {Minier}, {Molinari}, {Olofsson}, {Pezzuto}, {Russeil},
  {Roussel}, {Saraceno}, {Sauvage}, {Sibthorpe}, {Spinoglio}, {Testi},
  {Ward-Thompson}, {White}, {Wilson}, {Woodcraft}, \& {Zavagno}}]{konyves_2010}
{K{\"o}nyves}, V., {Andr{\'e}}, P., {Men'shchikov}, A., {et~al.} 2010, \aap,
  518, L106

\bibitem[{{Kroupa}(2001)}]{kroupa_2001}
{Kroupa}, P. 2001, \mnras, 322, 231

\bibitem[{{Lada} \& {Lada}(2003)}]{ladalada_2003}
{Lada}, C.~J. \& {Lada}, E.~A. 2003, \araa, 41, 57

\bibitem[{{Lada} {et~al.}(2010){Lada}, {Lombardi}, \& {Alves}}]{lada_2010}
{Lada}, C.~J., {Lombardi}, M., \& {Alves}, J.~F. 2010, \apj, 724, 687

\bibitem[{{Landy} \& {Szalay}(1993)}]{landy_1993}
{Landy}, S.~D. \& {Szalay}, A.~S. 1993, \apj, 412, 64

\bibitem[{{Li} {et~al.}(2016){Li}, {Burkert}, {Megeath}, \&
  {Wyrowski}}]{li_2016}
{Li}, G.-X., {Burkert}, A., {Megeath}, T., \& {Wyrowski}, F. 2016, ArXiv
  e-prints

\bibitem[{{Ling} {et~al.}(1986){Ling}, {Barrow}, \& {Frenk}}]{ling_1986}
{Ling}, E.~N., {Barrow}, J.~D., \& {Frenk}, C.~S. 1986, \mnras, 223, 21P

\bibitem[{{Mac Low} \& {Klessen}(2004)}]{mac-low_2004}
{Mac Low}, M.-M. \& {Klessen}, R.~S. 2004, Reviews of Modern Physics, 76, 125

\bibitem[{{Madden} {et~al.}(1997){Madden}, {Poglitsch}, {Geis}, {Stacey}, \&
  {Townes}}]{madden_1997}
{Madden}, S.~C., {Poglitsch}, A., {Geis}, N., {Stacey}, G.~J., \& {Townes},
  C.~H. 1997, \apj, 483, 200

\bibitem[{{Malinen} {et~al.}(2011){Malinen}, {Juvela}, {Collins}, {Lunttila},
  \& {Padoan}}]{malinen_2011}
{Malinen}, J., {Juvela}, M., {Collins}, D.~C., {Lunttila}, T., \& {Padoan}, P.
  2011, \aap, 530, A101

\bibitem[{{Meixner} {et~al.}(2006){Meixner}, {Gordon}, {Indebetouw}, {Hora},
  {Whitney}, {Blum}, {Reach}, {Bernard}, {Meade}, {Babler}, {Engelbracht},
  {For}, {Misselt}, {Vijh}, {Leitherer}, {Cohen}, {Churchwell}, {Boulanger},
  {Frogel}, {Fukui}, {Gallagher}, {Gorjian}, {Harris}, {Kelly}, {Kawamura},
  {Kim}, {Latter}, {Madden}, {Markwick-Kemper}, {Mizuno}, {Mizuno}, {Mould},
  {Nota}, {Oey}, {Olsen}, {Onishi}, {Paladini}, {Panagia}, {Perez-Gonzalez},
  {Shibai}, {Sato}, {Smith}, {Staveley-Smith}, {Tielens}, {Ueta}, {van Dyk},
  {Volk}, {Werner}, \& {Zaritsky}}]{meixner_2006}
{Meixner}, M., {Gordon}, K.~D., {Indebetouw}, R., {et~al.} 2006, \aj, 132, 2268

\bibitem[{{Meixner} {et~al.}(2013){Meixner}, {Panuzzo}, {Roman-Duval},
  {Engelbracht}, {Babler}, {Seale}, {Hony}, {Montiel}, {Sauvage}, {Gordon},
  {Misselt}, {Okumura}, {Chanial}, {Beck}, {Bernard}, {Bolatto}, {Bot},
  {Boyer}, {Carlson}, {Clayton}, {Chen}, {Cormier}, {Fukui}, {Galametz},
  {Galliano}, {Hora}, {Hughes}, {Indebetouw}, {Israel}, {Kawamura}, {Kemper},
  {Kim}, {Kwon}, {Lebouteiller}, {Li}, {Long}, {Madden}, {Matsuura}, {Muller},
  {Oliveira}, {Onishi}, {Otsuka}, {Paradis}, {Poglitsch}, {Reach},
  {Robitaille}, {Rubio}, {Sargent}, {Sewi{\l}o}, {Skibba}, {Smith},
  {Srinivasan}, {Tielens}, {van Loon}, \& {Whitney}}]{meixner_2013}
{Meixner}, M., {Panuzzo}, P., {Roman-Duval}, J., {et~al.} 2013, \aj, 146, 62

\bibitem[{{Motte} {et~al.}(2007){Motte}, {Bontemps}, {Schilke}, {Schneider},
  {Menten}, \& {Brogui{\`e}re}}]{motte_2007}
{Motte}, F., {Bontemps}, S., {Schilke}, P., {et~al.} 2007, \aap, 476, 1243

\bibitem[{{Mottram} {et~al.}(2011){Mottram}, {Hoare}, {Davies}, {Lumsden},
  {Oudmaijer}, {Urquhart}, {Moore}, {Cooper}, \& {Stead}}]{mottram_2011}
{Mottram}, J.~C., {Hoare}, M.~G., {Davies}, B., {et~al.} 2011, \apjl, 730, L33

\bibitem[{{Nayak} {et~al.}(2016){Nayak}, {Meixner}, {Indebetouw}, {De Marchi},
  {Koekemoer}, {Panagia}, \& {Sabbi}}]{nayak_2016}
{Nayak}, O., {Meixner}, M., {Indebetouw}, R., {et~al.} 2016, ArXiv e-prints

\bibitem[{{Pietrzy{\'n}ski} {et~al.}(2013){Pietrzy{\'n}ski}, {Graczyk},
  {Gieren}, {Thompson}, {Pilecki}, {Udalski}, {Soszy{\'n}ski}, {Koz{\l}owski},
  {Konorski}, {Suchomska}, {Bono}, {Moroni}, {Villanova}, {Nardetto},
  {Bresolin}, {Kudritzki}, {Storm}, {Gallenne}, {Smolec}, {Minniti}, {Kubiak},
  {Szyma{\'n}ski}, {Poleski}, {Wyrzykowski}, {Ulaczyk}, {Pietrukowicz},
  {G{\'o}rski}, \& {Karczmarek}}]{pietrzynski_2013}
{Pietrzy{\'n}ski}, G., {Graczyk}, D., {Gieren}, W., {et~al.} 2013, \nat, 495,
  76

\bibitem[{{Pineda} {et~al.}(2009){Pineda}, {Rosolowsky}, \&
  {Goodman}}]{pineda_2009}
{Pineda}, J.~E., {Rosolowsky}, E.~W., \& {Goodman}, A.~A. 2009, \apjl, 699,
  L134

\bibitem[{{Pon} {et~al.}(2011){Pon}, {Johnstone}, \& {Heitsch}}]{pon_2011}
{Pon}, A., {Johnstone}, D., \& {Heitsch}, F. 2011, \apj, 740, 88

\bibitem[{{Povich} {et~al.}(2009){Povich}, {Churchwell}, {Bieging}, {Kang},
  {Whitney}, {Brogan}, {Kulesa}, {Cohen}, {Babler}, {Indebetouw}, {Meade}, \&
  {Robitaille}}]{povich_2009}
{Povich}, M.~S., {Churchwell}, E., {Bieging}, J.~H., {et~al.} 2009, \apj, 696,
  1278

\bibitem[{{Raga} {et~al.}(2012){Raga}, {Cant{\'o}}, \&
  {Rodr{\'{\i}}guez}}]{raga_2012}
{Raga}, A.~C., {Cant{\'o}}, J., \& {Rodr{\'{\i}}guez}, L.~F. 2012, \mnras, 419,
  L39

\bibitem[{{Ragan} {et~al.}(2012){Ragan}, {Henning}, {Krause}, {Pitann},
  {Beuther}, {Linz}, {Tackenberg}, {Balog}, {Hennemann}, {Launhardt}, {Lippok},
  {Nielbock}, {Schmiedeke}, {Schuller}, {Steinacker}, {Stutz}, \&
  {Vasyunina}}]{ragan_2012}
{Ragan}, S., {Henning}, T., {Krause}, O., {et~al.} 2012, \aap, 547, A49

\bibitem[{{Ragan} {et~al.}(2014){Ragan}, {Henning}, {Tackenberg}, {Beuther},
  {Johnston}, {Kainulainen}, \& {Linz}}]{ragan_2014}
{Ragan}, S.~E., {Henning}, T., {Tackenberg}, J., {et~al.} 2014, \aap, 568, A73

\bibitem[{{Rathborne} {et~al.}(2006){Rathborne}, {Jackson}, \&
  {Simon}}]{rathborne_2006}
{Rathborne}, J.~M., {Jackson}, J.~M., \& {Simon}, R. 2006, \apj, 641, 389

\bibitem[{{Roberts}(1969)}]{roberts_1969}
{Roberts}, W.~W. 1969, \apj, 158, 123

\bibitem[{{Robitaille}(2008)}]{robitaille_2008}
{Robitaille}, T.~P. 2008, in Astronomical Society of the Pacific Conference
  Series, Vol. 387, Massive Star Formation: Observations Confront Theory, ed.
  H.~{Beuther}, H.~{Linz}, \& T.~{Henning}, 290

\bibitem[{{Robitaille} {et~al.}(2006){Robitaille}, {Whitney}, {Indebetouw},
  {Wood}, \& {Denzmore}}]{robitaille_2006}
{Robitaille}, T.~P., {Whitney}, B.~A., {Indebetouw}, R., {Wood}, K., \&
  {Denzmore}, P. 2006, \apjs, 167, 256

\bibitem[{{Roman-Duval} {et~al.}(2014){Roman-Duval}, {Gordon}, {Meixner},
  {Bot}, {Bolatto}, {Hughes}, {Wong}, {Babler}, {Bernard}, {Clayton}, {Fukui},
  {Galametz}, {Galliano}, {Glover}, {Hony}, {Israel}, {Jameson},
  {Lebouteiller}, {Lee}, {Li}, {Madden}, {Misselt}, {Montiel}, {Okumura},
  {Onishi}, {Panuzzo}, {Reach}, {Remy-Ruyer}, {Robitaille}, {Rubio}, {Sauvage},
  {Seale}, {Sewilo}, {Staveley-Smith}, \& {Zhukovska}}]{roman-duval_2014}
{Roman-Duval}, J., {Gordon}, K.~D., {Meixner}, M., {et~al.} 2014, \apj, 797, 86

\bibitem[{{Rosolowsky} {et~al.}(2008){Rosolowsky}, {Pineda}, {Kauffmann}, \&
  {Goodman}}]{rosolowsky_2008}
{Rosolowsky}, E.~W., {Pineda}, J.~E., {Kauffmann}, J., \& {Goodman}, A.~A.
  2008, \apj, 679, 1338

\bibitem[{{Schneider} {et~al.}(2012){Schneider}, {Csengeri}, {Hennemann},
  {Motte}, {Didelon}, {Federrath}, {Bontemps}, {Di Francesco}, {Arzoumanian},
  {Minier}, {Andr{\'e}}, {Hill}, {Zavagno}, {Nguyen-Luong}, {Attard},
  {Bernard}, {Elia}, {Fallscheer}, {Griffin}, {Kirk}, {Klessen}, {K{\"o}nyves},
  {Martin}, {Men'shchikov}, {Palmeirim}, {Peretto}, {Pestalozzi}, {Russeil},
  {Sadavoy}, {Sousbie}, {Testi}, {Tremblin}, {Ward-Thompson}, \&
  {White}}]{schneider_2012}
{Schneider}, N., {Csengeri}, T., {Hennemann}, M., {et~al.} 2012, \aap, 540, L11

\bibitem[{{Scoville} {et~al.}(1986){Scoville}, {Sanders}, \&
  {Clemens}}]{scoville_1986}
{Scoville}, N.~Z., {Sanders}, D.~B., \& {Clemens}, D.~P. 1986, \apjl, 310, L77

\bibitem[{{Scoville} {et~al.}(1987){Scoville}, {Yun}, {Sanders}, {Clemens}, \&
  {Waller}}]{scoville_1987}
{Scoville}, N.~Z., {Yun}, M.~S., {Sanders}, D.~B., {Clemens}, D.~P., \&
  {Waller}, W.~H. 1987, \apjs, 63, 821

\bibitem[{{Seale} {et~al.}(2012){Seale}, {Looney}, {Wong}, {Ott}, {Klein}, \&
  {Pineda}}]{seale_2012}
{Seale}, J.~P., {Looney}, L.~W., {Wong}, T., {et~al.} 2012, \apj, 751, 42

\bibitem[{{Seale} {et~al.}(2014){Seale}, {Meixner}, {Sewi{\l}o}, {Babler},
  {Engelbracht}, {Gordon}, {Hony}, {Misselt}, {Montiel}, {Okumura}, {Panuzzo},
  {Roman-Duval}, {Sauvage}, {Boyer}, {Chen}, {Indebetouw}, {Matsuura},
  {Oliveira}, {Srinivasan}, {van Loon}, {Whitney}, \& {Woods}}]{seale_2014}
{Seale}, J.~P., {Meixner}, M., {Sewi{\l}o}, M., {et~al.} 2014, \aj, 148, 124

\bibitem[{{Shetty} {et~al.}(2009){Shetty}, {Kauffmann}, {Schnee}, \&
  {Goodman}}]{shetty_2009}
{Shetty}, R., {Kauffmann}, J., {Schnee}, S., \& {Goodman}, A.~A. 2009, \apj,
  696, 676

\bibitem[{{Shu} {et~al.}(1987){Shu}, {Adams}, \& {Lizano}}]{shu_1987}
{Shu}, F.~H., {Adams}, F.~C., \& {Lizano}, S. 1987, \araa, 25, 23

\bibitem[{{Simon} {et~al.}(2006){Simon}, {Rathborne}, {Shah}, {Jackson}, \&
  {Chambers}}]{simon_2006}
{Simon}, R., {Rathborne}, J.~M., {Shah}, R.~Y., {Jackson}, J.~M., \&
  {Chambers}, E.~T. 2006, \apj, 653, 1325

\bibitem[{{Smith} \& {MCELS Team}(1998)}]{smith_1998}
{Smith}, R.~C. \& {MCELS Team}. 1998, \pasa, 15, 163

\bibitem[{{Spitzer}(1978)}]{spitzer_1978}
{Spitzer}, L. 1978, {Physical processes in the interstellar medium}

\bibitem[{{Svoboda} {et~al.}(2015){Svoboda}, {Shirley}, {Battersby},
  {Rosolowsky}, {Ginsburg}, {Ellsworth-Bowers}, {Pestalozzi}, {Dunham},
  {Evans}, {Bally}, \& {Glenn}}]{svoboda_2015}
{Svoboda}, B.~E., {Shirley}, Y.~L., {Battersby}, C., {et~al.} 2015, ArXiv
  e-prints

\bibitem[{{Tackenberg} {et~al.}(2012){Tackenberg}, {Beuther}, {Henning},
  {Schuller}, {Wienen}, {Motte}, {Wyrowski}, {Bontemps}, {Bronfman}, {Menten},
  {Testi}, \& {Lefloch}}]{tackenberg_2012}
{Tackenberg}, J., {Beuther}, H., {Henning}, T., {et~al.} 2012, \aap, 540, A113

\bibitem[{{Tan} {et~al.}(2014){Tan}, {Beltr{\'a}n}, {Caselli}, {Fontani},
  {Fuente}, {Krumholz}, {McKee}, \& {Stolte}}]{tan_2014}
{Tan}, J.~C., {Beltr{\'a}n}, M.~T., {Caselli}, P., {et~al.} 2014, Protostars
  and Planets VI, 149

\bibitem[{{Tenorio-Tagle} \& {Bodenheimer}(1988)}]{tenorio-tagle_1988}
{Tenorio-Tagle}, G. \& {Bodenheimer}, P. 1988, \araa, 26, 145

\bibitem[{{Thompson} {et~al.}(2012){Thompson}, {Urquhart}, {Moore}, \&
  {Morgan}}]{thompson_2012}
{Thompson}, M.~A., {Urquhart}, J.~S., {Moore}, T.~J.~T., \& {Morgan}, L.~K.
  2012, \mnras, 421, 408

\bibitem[{{Tout} {et~al.}(1996){Tout}, {Pols}, {Eggleton}, \&
  {Han}}]{tout_1996}
{Tout}, C.~A., {Pols}, O.~R., {Eggleton}, P.~P., \& {Han}, Z. 1996, \mnras,
  281, 257

\bibitem[{{Tremblin} {et~al.}(2014){Tremblin}, {Anderson}, {Didelon}, {Raga},
  {Minier}, {Ntormousi}, {Pettitt}, {Pinto}, {Samal}, {Schneider}, \&
  {Zavagno}}]{tremblin_2014}
{Tremblin}, P., {Anderson}, L.~D., {Didelon}, P., {et~al.} 2014, \aap, 568, A4

\bibitem[{{Whitney} {et~al.}(2008){Whitney}, {Sewilo}, {Indebetouw},
  {Robitaille}, {Meixner}, {Gordon}, {Meade}, {Babler}, {Harris}, {Hora},
  {Bracker}, {Povich}, {Churchwell}, {Engelbracht}, {For}, {Block}, {Misselt},
  {Vijh}, {Leitherer}, {Kawamura}, {Blum}, {Cohen}, {Fukui}, {Mizuno},
  {Mizuno}, {Srinivasan}, {Tielens}, {Volk}, {Bernard}, {Boulanger}, {Frogel},
  {Gallagher}, {Gorjian}, {Kelly}, {Latter}, {Madden}, {Kemper}, {Mould},
  {Nota}, {Oey}, {Olsen}, {Onishi}, {Paladini}, {Panagia}, {Perez-Gonzalez},
  {Reach}, {Shibai}, {Sato}, {Smith}, {Staveley-Smith}, {Ueta}, {Van Dyk},
  {Werner}, {Wolff}, \& {Zaritsky}}]{whitney_2008}
{Whitney}, B.~A., {Sewilo}, M., {Indebetouw}, R., {et~al.} 2008, \aj, 136, 18

\bibitem[{{Williams} {et~al.}(1994){Williams}, {de Geus}, \&
  {Blitz}}]{williams_1994}
{Williams}, J.~P., {de Geus}, E.~J., \& {Blitz}, L. 1994, \apj, 428, 693

\bibitem[{{Wilson} {et~al.}(2005){Wilson}, {Dame}, {Masheder}, \&
  {Thaddeus}}]{wilson_2005}
{Wilson}, B.~A., {Dame}, T.~M., {Masheder}, M.~R.~W., \& {Thaddeus}, P. 2005,
  \aap, 430, 523

\bibitem[{{Wong} {et~al.}(2011){Wong}, {Hughes}, {Ott}, {Muller}, {Pineda},
  {Bernard}, {Chu}, {Fukui}, {Gruendl}, {Henkel}, {Kawamura}, {Klein},
  {Looney}, {Maddison}, {Mizuno}, {Paradis}, {Seale}, \& {Welty}}]{wong_2011}
{Wong}, T., {Hughes}, A., {Ott}, J., {et~al.} 2011, \apjs, 197, 16

\bibitem[{{Woods} {et~al.}(2006){Woods}, {Geller}, \& {Barton}}]{woods_2006}
{Woods}, D.~F., {Geller}, M.~J., \& {Barton}, E.~J. 2006, \aj, 132, 197

\bibitem[{{Ysard} {et~al.}(2012){Ysard}, {Juvela}, {Demyk}, {Guillet},
  {Abergel}, {Bernard}, {Malinen}, {M{\'e}ny}, {Montier}, {Paradis},
  {Ristorcelli}, \& {Verstraete}}]{ysard_2012}
{Ysard}, N., {Juvela}, M., {Demyk}, K., {et~al.} 2012, \aap, 542, A21

\bibitem[{{Zavagno} {et~al.}(2007){Zavagno}, {Pomar{\`e}s}, {Deharveng},
  {Hosokawa}, {Russeil}, \& {Caplan}}]{zavagno_2007}
{Zavagno}, A., {Pomar{\`e}s}, M., {Deharveng}, L., {et~al.} 2007, \aap, 472,
  835

\bibitem[{{Zavagno} {et~al.}(2010){Zavagno}, {Russeil}, {Motte}, {Anderson},
  {Deharveng}, {Rod{\'o}n}, {Bontemps}, {Abergel}, {Baluteau}, {Sauvage},
  {Andr{\'e}}, {Hill}, \& {White}}]{zavagno_2010}
{Zavagno}, A., {Russeil}, D., {Motte}, F., {et~al.} 2010, \aap, 518, L81

\end{thebibliography}
\end{document}